%% file: Nskin-revised.tex
\documentclass[utf8]{FrontiersinHarvard}

\usepackage{url,hyperref,lineno,microtype,subcaption}
\usepackage[onehalfspacing]{setspace}
\usepackage{bm}
\usepackage{booktabs}
\usepackage{braket}
\usepackage{lscape}

\def\ColorA{black}
\def\ColorB{black}

\newcommand{\RevA}[1]{{\color{\ColorA}{#1}}}
\newcommand{\RevB}[1]{{\color{\ColorB}{#1}}}

\def\keyFont{\fontsize{8}{11}\helveticabold }
\def\firstAuthorLast{Miyatsu {et~al.}} %use et al only if is more than 1 author
\def\Authors{Tsuyoshi Miyatsu\,$^{1,*}$, Myung-Ki Cheoun\,$^{1}$, Kyungsik Kim\,$^{2}$ and Koichi Saito\,$^{3}$}
\def\Address{
  $^{1}$Department of Physics and OMEG Institute, Soongsil University, Seoul 06978, Republic of Korea \\
  $^{2}$School of Liberal Arts and Sciences, Korea Aerospace University, Goyang 10540, Republic of Korea \\
  $^{3}$Department of Physics and Astronomy, Faculty of Science and Technology, Tokyo University of Science, Noda 278-8510, Japan
}
\def\corrAuthor{Tsuyoshi Miyatsu}

\def\corrEmail{tsuyoshi.miyatsu@ssu.ac.kr}

\begin{document}
\onecolumn
\firstpage{1}

\title[Novel features of asymmetric nuclear matter]
{
  Novel features of asymmetric nuclear matter %from large neutron skin thickness and small neutron-star radii
  \RevB{from terrestrial experiments and astrophysical observations of neutron stars}
}

\author[\firstAuthorLast ]{\Authors} %This field will be automatically populated
\address{} %This field will be automatically populated
\correspondance{} %This field will be automatically populated

\extraAuth{}% If there are more than 1 corresponding author, comment this line and uncomment the next one.
%\extraAuth{corresponding Author2 \\ Laboratory X2, Institute X2, Department X2, Organization X2, Street X2, City X2 , State XX2 (only USA, Canada and Australia), Zip Code2, X2 Country X2, email2@uni2.edu}

\maketitle

\begin{abstract}
  The accurate measurement of neutron skin thickness of $^{208}$Pb by the PREX Collaboration suggests a large value of the nuclear symmetry energy slope parameter, $L$, whereas the smaller $L$ is preferred to account for the small neutron-star radii from NICER observations.
  To resolve this discrepancy between nuclear experiments and astrophysical observations, new effective interactions have been developed using relativistic mean-field models with the isoscalar- and isovector-meson mixing.
  We investigate the effects of $\delta$-nucleon coupling and $\sigma$--$\delta$ mixing on the ground-state properties of finite nuclei, as well as the characteristics of isospin-asymmetric nuclear matter and neutron stars.
  Additionally, we explore the role of the quartic $\rho$-meson self-interaction in dense nuclear matter to mitigate the stiff equation of state for neutron stars resulting from the large $\delta$-nucleon coupling.
  It is found that the nuclear symmetry energy undergoes a sudden softening at approximately twice the saturation density of nuclear matter, taking into account the PREX-2 result, the recent NICER observation of PSR J0437$–$4715, and the binary neutron star merger, GW170817.

%%% Leave the Abstract empty if your article does not require one, please see the Summary Table for full details.
% \section{}
% For full guidelines regarding your manuscript please refer to \href{http://www.frontiersin.org/about/AuthorGuidelines}{Author Guidelines}.

% As a primary goal, the abstract should render the general significance and conceptual advance of the work clearly accessible to a broad readership. References should not be cited in the abstract. Leave the Abstract empty if your article does not require one, please see \href{http://www.frontiersin.org/about/AuthorGuidelines#SummaryTable}{Summary Table} for details according to article type.

\tiny
\keyFont{ \section{Keywords:}
  isospin-asymmetric nuclear matter,
  neutron skin thickness,
  neutron stars,
  NICER,
  nuclear equation of state,
  nuclear symmetry energy,
  PREX-2,
  relativistic mean-field models
} %All article types: you may provide up to 8 keywords; at least 5 are mandatory.
\end{abstract}

\section{Introduction} \label{sec:introduction}

The astrophysical phenomena concerning compact stars as well as the characteristics of finite nuclei and nuclear matter are determined by the nuclear equation of state (EoS), characterized by the relation between the energy density and pressure of the system~\citep{Baym:1971ax,Bethe:1979zd}.
Many nuclear EoSs have been contemplated so far through realistic nuclear models in a nonrelativistic or relativistic framework~\citep{Lattimer:1991nc,Glendenning:1991es}.
Relativistic mean-field (RMF) calculations, based on the one-boson exchange potential for nuclear interactions~\citep{Machleidt:1987hj,Machleidt:1989tm}, have achieved great success in understanding of the properties of nuclear matter and finite nuclei~\citep{Serot:1984ey}.
\RevB{To reproduce a reasonable nuclear incompressibility and properties of unstable nuclei, the RMF models have been developed by introducing the nonlinear self-couplings of isoscalar, Lorentz-scalar ($\sigma$) and Lorentz-vector ($\omega^{\mu}$) mesons~\citep{Boguta:1977xi,Sugahara:1993wz}. In addition, the isovector, Lorentz-vector ($\bm{\rho}^{\mu}$) meson and its nonlinear couplings have been considered to describe a neutron skin thickness of heavy nuclei and characteristics of isospin-asymmetric nuclear matter~\citep{Mueller:1996pm,Horowitz:2001ya}.}
The RMF approach is, at present, one of the most powerful tools to study neutron star physics~\citep{Oertel:2016bki,Alford:2022bpp,Patra:2023jvv}, as in the case of the Skyrme energy density functional~\citep{Stone:2021uju,Stone:2024inj,Zhou:2023hzu,Sun:2023xkg}.

The nuclear symmetry energy, $E_{\rm sym}$, which is defined as the difference between the energies of pure neutron and symmetric nuclear matter, is recognized to be an important physical quantity to study the properties of isospin-asymmetric nuclear EoS~\citep{Li:2008gp,Lattimer:2014scr}.
In addition, the slope parameter of nuclear symmetry energy, $L$, gives a significant constraint on the density dependence of $E_{\rm sym}$ and is related to the neutron skin thickness of heavy nuclei~\citep{Typel:2001lcw}.
Laboratory experiments have been also performed to investigate the properties of low-density nuclear matter and to impose constraints on $E_{\rm sym}$ and $L$ through the heavy-ion collisions (HICs)~\citep{Tsang:2008fd,Tsang:2012se}.
Recently, the impacts of the higher-order coefficients---the curvature and skewness of nuclear symmetry energy, $K_{\rm sym}$ and $J_{\rm sym}$---have been studied in light of some astrophysical observations, for instance the mass-radius relations of neutron stars and the cooling process of proto-neutron stars~\citep{Zhang:2021xdt,Richter:2023zec,Xie:2024mxu}.

Owing to the precise observations of neutron stars, such as the Shapiro delay measurement of a binary millisecond pulsar J1614$-$2230~\citep{Demorest:2010bx,NANOGrav:2017wvv} and the radius measurement of PSR J0740$+$6620 from Neutron Star Interior Composition Explorer (NICER) and from X-ray Multi-Mirror (XMM-Newton) Data~\citep{Miller:2019cac,Miller:2021qha,Riley:2019yda,Riley:2021pdl}, theoretical studies have been currently performed more than ever to elucidate neutron star physics through the nuclear EoS for dense matter.
It has been found that the nuclear EoS should satisfy at least $2M_{\odot}$ to support the high-mass PSR J0740$+$6620 event, and that the precise measurements of neutron-star radii provide the valuable information in determining the features of isospin-asymmetric nuclear matter.
In addition, the direct detection of gravitational-wave (GW) signals from a binary neutron star merger, GW170817, observed by Advanced LIGO and Advanced Virgo detectors has placed stringent restrictions on the mass--radius relation of neutron stars~\citep{LIGOScientific:2017vwq,LIGOScientific:2018cki,LIGOScientific:2018hze}.
In particular, the tidal deformability of a neutron star~\citep{Hinderer:2007mb,Hinderer:2009ca} plays a critical role in constructing the EoS for neutron star matter~\citep{Annala:2017llu,Lim:2018bkq,Most:2018hfd,Raithel:2018ncd}.
It has been reported that there are the strong correlations of neutron-star radii with $E_{\rm sym}$ and $L$, and the radius of a typical neutron star is determined by $L$~\citep{Alam:2016cli,Hu:2020ujf,Li:2020ass,Lopes:2024bvz}.
Using a Bayesian analysis based on constraints from NICER and GW170817 within chiral effective field theory calculations, $L$ is currently estimated as $L=(43.7$--$70.0)$~MeV~\citep{Lim:2023dbk}.

The accurate measurement of neutron skin thickness of $^{208}$Pb, $R_{\rm skin}^{208}$, by the PREX Collaboration, using the parity-violating electron scattering, has revealed a serious discrepancy between the measured $R_{\rm skin}^{208}$ and theoretical predictions~\citep{PREX:2021umo}.
\RevA{
  The neutron skin thickness, $R_{\rm skin}$, is defined here as the difference between the root-mean-square radii of point neutrons and protons, $R_{n}$ and $R_{p}$, in a nucleus:
  \begin{equation}
    R_{\rm skin}=R_{n}-R_{p}.
  \end{equation}
}To explain the PREX-2 result, \citet{Reed:2021nqk} have proposed the large $L$ value as $L=106\pm37$~MeV, by exploiting the strong correlation between $R_{\rm skin}^{208}$ and $L$.
In contrast, \citet{Reinhard:2021utv}, using modern relativistic and nonrelativistic energy density functionals, have predicted the smaller value, $L=54\pm8$~MeV, by carefully assessing theoretical uncertainty on the parity-violating asymmetry, $A_{\rm PV}$, in $^{208}$Pb.
Additionally, the CREX experiment, which provides a precise measurement of the neutron skin thickness of $^{48}$Ca, $R_{\rm skin}^{48}$, through the parity-violating electron scattering~\citep{CREX:2022kgg}, complicates the understanding of isospin-asymmetric nuclear matter.
This complexity arises from the difficulty of reconciling the PREX-2 and CREX results simultaneously.
\RevB{In addition, the measurements from polarized proton scattering off $^{208}$Pb indicate smaller $R_{\rm skin}^{208}$,  and consequently smaller $L$, compared to those obtained from the PREX-2 experiment~\citep{Zenihiro:2010zz,Tamii:2011pv}. As a result, $R_{\rm skin}^{208}$ and $L$ remain uncertain in theoretical calculations~\citep{Centelles:2008vu,Dutra:2014qga}.  At present, many species of neutron skin thickness have been reported from a combination of experimental and theoretical results~\citep{Trzcinska:2001sy}.}

In this article, we review the recently updated RMF models with nonlinear couplings by introducing the isoscalar- and isovector-meson mixing, $\sigma^{2}\bm{\delta}^{2}$ and $\omega_{\mu}\omega^{\mu}\bm{\rho}_{\nu}\bm{\rho}^{\nu}$, which can cover both data from stable nuclear ground states and astrophysical observations of neutron stars.
Although the isovector, Lorentz-scalar ($\bm{\delta}$) meson has been claimed to be less important than the isovector, Lorentz-vector ($\bm{\rho}^{\mu}$) meson so far, it has been recently realized that the $\delta$ meson considerably affects the properties of isospin-asymmetric nuclear EoS, such as neutron skin thickness of heavy nuclei and neutron-star radii~\citep{Zabari:2018tjk,Miyatsu:2022wuy,Miyatsu:2023lki,Li:2022okx}
The new effective interactions discussed in this review are constructed under the constraints from the terrestrial experiments and astrophysical observations of neutron stars, especially focusing on the PREX-2 and CREX experiments.
The resulting nuclear EoS have to support the following conditions:
\begin{itemize}
\item[(1)] the EoSs for symmetric nuclear matter and pure neutron matter satisfy the particle flow data in heavy-ion collisions (HICs)~\citep{Danielewicz:2002pu,Fuchs:2005zg,Lynch:2009vc,Oliinychenko:2022uvy},
\item[(2)] the EoS for neutron stars attains to the observed mass of PSR J0740$+$6620 ($M=2.072^{+0.067}_{-0.066}$~$M_{\odot}$)~\citep{NANOGrav:2019jur,Fonseca:2021wxt,Riley:2021pdl},
\item[(3)] the EoS for neutron stars explains the dimensionless tidal deformability from the binary merger event, GW170817 ($\Lambda_{1.4}=190^{+390}_{-120}$)~\citep{LIGOScientific:2018cki,LIGOScientific:2018hze}.
\end{itemize}
Under these constraints, we examine the effects of the $\delta$-nucleon coupling and $\sigma$--$\delta$ mixing on the ground-state properties of finite nuclei, and consider the PREX-2 and CREX results.
Additionally, we investigate the impact of the quartic self-interactions of $\delta$ and $\rho$ mesons on the nuclear EoS to study the properties of neutron star matter.

This paper is organized as follows.
A summary and analytical calculations concerning the RMF model with nonlinear couplings are described in Sec.~\ref{sec:framework}.
Numerical results and detailed discussions are presented in Sec.~\ref{sec:results}.
Finally, we give a summary in Sec.~\ref{sec:summary}.

% \clearpage

\section{Theoretical framework} \label{sec:framework}

\subsection{Lagrangian density}

In quantum hadrodynamics~\citep{Serot:1984ey}, we employ the recently updated effective Lagrangian density including the isoscalar ($\sigma$ and $\omega^{\mu}$) and isovector ($\bm{\delta}$ and $\bm{\rho}^{\mu}$) mesons as well as nucleons ($N=p,n$)~\citep{Miyatsu:2022wuy,Miyatsu:2023lki}.
The total Lagrangian density is then given by
\begin{align}
  \mathcal{L}
  & = \bar{\psi}_{N} \left[ i\gamma_{\mu}\partial^{\mu}
    - \left( M_{N} - g_{\sigma}\sigma - g_{\delta}\bm{\delta}\cdot\bm{\tau}_{N} \right)
    - g_{\omega}\gamma_{\mu}\omega^{\mu} - g_{\rho}\gamma_{\mu}\bm{\rho}^{\mu}\cdot\bm{\tau}_{N} \right]\psi_{N}
    \nonumber \\
  & + \frac{1}{2}\left(\partial_{\mu}\sigma\partial^{\mu}\sigma-m_{\sigma}^{2}\sigma^{2}\right)
    + \frac{1}{2}m_{\omega}^{2}\omega_{\mu}\omega^{\mu}-\frac{1}{4}W_{\mu\nu}W^{\mu\nu}
    + \frac{1}{2}\left(\partial_{\mu}\bm{\delta}\cdot\partial^{\mu}\bm{\delta}-m_{\delta}^{2}\bm{\delta}\cdot\bm{\delta}\right)
    \nonumber \\
  & + \frac{1}{2}m_{\rho}^{2}\bm{\rho}_{\mu}\cdot\bm{\rho}^{\mu}-\frac{1}{4}\bm{R}_{\mu\nu}\cdot\bm{R}^{\mu\nu}
    + \mathcal{L}_{\rm EM}
    - U_{\rm NL}(\sigma,\omega,\bm{\delta},\bm{\rho}),
    \label{eq:Ltot}
\end{align}
where $\psi_{N}={\psi_{p} \choose \psi_{n}}$ is the iso-doublet, nucleon field, $\bm{\tau}_{N}$ is its isospin matrix, $W_{\mu\nu}=\partial_{\mu}\omega_{\nu}-\partial_{\nu}\omega_{\mu}$, and $\bm{R}_{\mu\nu}=\partial_{\mu}\bm{\rho}_{\nu}-\partial_{\nu}\bm{\rho}_{\mu}$.
The meson-nucleon coupling constants are respectively denoted by $g_{\sigma}$, $g_{\omega}$, $g_{\delta}$, and $g_{\rho}$.
The photon-$N$ interaction, $\mathcal{L}_{\rm EM}=-e\bar{\psi}_{p}\gamma_{\mu}A^{\mu}\psi_{p}-\frac{1}{4}F_{\mu\nu}F^{\mu\nu}$ with $F_{\mu\nu}=\partial_{\mu}A_{\nu}-\partial_{\nu}A_{\mu}$, is also taken into account to describe the characteristics of finite nuclei~\citep{Serot:1984ey,Ring:1996qi}.
Additionally, a nonlinear potential in Eq.~\eqref{eq:Ltot} is supplemented as follows:
\begin{align}
  U_{\rm NL}(\sigma,\omega,\bm{\delta},\bm{\rho})
  & = \frac{1}{3}g_{2}\sigma^{3} + \frac{1}{4}g_{3}\sigma^{4} - \frac{1}{4}c_{3}(\omega_{\mu}\omega^{\mu})^{2}
    + \frac{1}{4}d_{3}(\bm{\delta}\cdot\bm{\delta})^{2} - \frac{1}{4}e_{3}(\bm{\rho}_{\mu}\cdot\bm{\rho}^{\mu})^{2}
    \nonumber \\
  & - \Gamma_{\sigma\delta}\sigma(\bm{\delta}\cdot\bm{\delta})
    - \Lambda_{\sigma\delta}\sigma^{2}(\bm{\delta}\cdot\bm{\delta})
    - \Lambda_{\omega\rho}(\omega_{\mu}\omega^{\mu})(\bm{\rho}_{\nu}\cdot\bm{\rho}^{\nu}).
    \label{eq:NL-pot}
\end{align}
The first and second terms in Eq.~\eqref{eq:NL-pot} are introduced to obtain a quantitative description of ground-state properties for symmetric nuclear matter~\citep{Boguta:1977xi,Lalazissis:1996rd}.
The quartic self-interactions of $\omega$, $\delta$, and $\rho$ mesons are also introduced in Eq.~\eqref{eq:NL-pot}~\citep{Sugahara:1993wz,Mueller:1996pm,Pradhan:2022txg,Malik:2024qjw}.
We also consider the isoscalar- and isovector-meson mixing, which only affects the characteristics of $N\not=Z$ finite nuclei and isospin-asymmetric nuclear matter~\citep{Todd-Rutel:2005yzo,Miyatsu:2013yta,Zabari:2018tjk}, while the scalar-vector mixing is not included in the present study~\citep{Horowitz:2000xj,Horowitz:2001ya,Haidari:2007cg,Sharma:2008jza,Kubis:2023gxa}.

\subsection{Field equations for finite nuclei in mean-field approximation}

In mean-field approximation, the meson and photon fields are replaced by the mean-field values: $\bar{\sigma}$, $\bar{\omega}$, $\bar{\delta}$, $\bar{\rho}$, and $\bar{A}$.
Then, the effective nucleon mass in matter is simply expressed as
\begin{equation}
  M_{N={p \choose n}}^{\ast}(\bar{\sigma},\bar{\delta}) = M_{N} - g_{\sigma}\bar{\sigma} \mp g_{\delta}\bar{\delta},
  \label{eq:emass}
\end{equation}
where $M_{N}$~($=939$~MeV) is the nucleon mass in free space.
If we restrict consideration to spherical finite nuclei, the equation of motion for $N$ is given by
\begin{equation}
  \left[-i\bm{\alpha}\cdot\bm{\nabla}+\beta M_{p \choose n}^{\ast}(\bar{\sigma},\bar{\delta})
    + g_{\omega}\bar{\omega} \pm g_{\rho}\bar{\rho} + e\frac{1\pm1}{2}\bar{A} \right]\,\psi_{p \choose n}
  = E_{\alpha{p \choose n}}\psi_{p \choose n},
  \label{eq:Dirac}
\end{equation}
with $E_{\alpha N}$ being the nucleon single-particle energy.
The meson and photon fields are then given by
\begin{align}
  \left[-\bm{\nabla}^{2}+m_{\sigma}^{\ast2}(\bar{\sigma},\bar{\delta})\right]\bar{\sigma}
  & = g_{\sigma}\left(\rho_{p}^{s}+\rho_{n}^{s}\right),
    \label{eq:sigma} \\
  \left[-\bm{\nabla}^{2}+m_{\omega}^{\ast2}(\bar{\omega},\bar{\rho})\right]\bar{\omega}
  & = g_{\omega}\left(\rho_{p}+\rho_{n}\right),
    \label{eq:omega} \\
  \left[-\bm{\nabla}^{2}+m_{\delta}^{\ast2}(\bar{\sigma},\bar{\delta})\right]\bar{\delta}
  & = g_{\delta}\left(\rho_{p}^{s}-\rho_{n}^{s}\right),
    \label{eq:delta} \\
  \left[-\bm{\nabla}^{2}+m_{\rho}^{\ast2}(\bar{\omega},\bar{\rho})\right]\bar{\rho}
  & = g_{\rho}\left(\rho_{p}-\rho_{n}\right),
    \label{eq:rho} %\\
%  -\bm{\nabla}^{2}\bar{A}
%  & = e\rho_{p},
%    \label{eq:photon}
\end{align}
and
\begin{equation}
  -\bm{\nabla}^{2}\bar{A} = e\rho_{p},
  \label{eq:photon}
\end{equation}
where $\rho_{N}^{s}$ ($\rho_{N}$) is the scalar (baryon) density for $N$, \RevA{which is computed self-consistently using nucleon wave functions in Eq.~\eqref{eq:Dirac} that are solutions to the Dirac equation in the spatially dependent meson and photon fields.}
The effective meson masses are defined by
\begin{align}
  m_{\sigma}^{\ast2}(\bar{\sigma},\bar{\delta})
  & = m_{\sigma}^{2} + g_{2}\bar{\sigma} + g_{3}\bar{\sigma}^{2}
    - \Gamma_{\sigma\delta}\bar{\delta}^{2}/\bar{\sigma} - 2\Lambda_{\sigma\delta}\bar{\delta}^{2},
    \label{eq:sigma-mass} \\
  m_{\omega}^{\ast2}(\bar{\omega},\bar{\rho})
  & = m_{\omega}^{2} + c_{3}\bar{\omega}^{2} + 2\Lambda_{\omega\rho}\bar{\rho}^{2},
    \label{eq:omega-mass} \\
  m_{\delta}^{\ast2}(\bar{\sigma},\bar{\delta})
  & = m_{\delta}^{2} + d_{3} \bar{\delta}^{2} - 2\Gamma_{\sigma\delta}\bar{\sigma} - 2\Lambda_{\sigma\delta}\bar{\sigma}^{2},
    \label{eq:delta-mass} \\
  m_{\rho}^{\ast2}(\bar{\omega},\bar{\rho})
  & = m_{\rho}^{2} + e_{3}\bar{\rho}^{2} + 2\Lambda_{\omega\rho}\bar{\omega}^{2}.
    \label{eq:rho-mass}
\end{align}
The total energy of the system is thus written as
\begin{align}
  E_{\rm tot}
  & = \sum_{N=p,n}\sum_{\alpha}^{\rm occ}\left(2j_{\alpha}+1\right)E_{\alpha N}
    \nonumber \\
  & + \frac{1}{2}\int d\bm{r} \,\left[g_{\sigma}\left(\rho_{p}^{s}+\rho_{n}^{s}\right)\bar{\sigma}
    - g_{\omega}\left(\rho_{p}+\rho_{n}\right)\bar{\omega}
    + g_{\delta}\left(\rho_{p}^{s}-\rho_{n}^{s}\right)\bar{\delta}
    - g_{\rho}\left(\rho_{p}-\rho_{n}\right)\bar{\rho}-e\rho_{p}\bar{A} \right]
    \nonumber \\
  & + \frac{1}{2}\int d\bm{r} \,\left( - \frac{1}{3}g_{2}\bar{\sigma}^{3} - \frac{1}{2}g_{3}\bar{\sigma}^{4}
    + \frac{1}{2}c_{3}\bar{\omega}^{4} - \frac{1}{2}d_{3}\bar{\delta}^{4} + \frac{1}{2}e_{3}\bar{\rho}^{4}
    + \Gamma_{\sigma\delta}\bar{\sigma}\bar{\delta}^{2} + 2\Lambda_{\sigma\delta}\bar{\sigma}^{2}\bar{\delta}^{2}
    + 2\Lambda_{\omega\rho}\bar{\omega}^{2}\bar{\rho}^{2} \right),
\end{align}
where the sum $\alpha$ runs over the occupied states of $E_{\alpha N}$ with the degeneracy $\left(2j_{\alpha}+1\right)$~\citep{Serot:1984ey}.

\subsection{Infinite nuclear matter}

To study the bulk properties of nuclear and neutron star matter, it is necessary to compute the nuclear equation of state (EoS)---a relation between the energy density, $\varepsilon_{B}$, and pressure, $P_{B}$.
In infinite nuclear matter, the surface terms in Eqs.~\eqref{eq:sigma}--\eqref{eq:photon} have no influence on its characteristics as the gradient reads zero.
The scalar and baryon density for $N~\left(=p,n\right)$ are then obtained as
\begin{align}
  \rho_{N}^{s}
  & = \braket{\bar{\psi}_{N}\psi_{N}}
    \nonumber \\
  & = \frac{1}{\pi^{2}} \int_{0}^{k_{F_{N}}} dk \, k^{2} \frac{M_{N}^{\ast}}{\sqrt{k^{2}+M_{N}^{\ast2}}}
    = \frac{M_{N}^{\ast3}}{2\pi^{2}} \left[ \frac{k_{F_{N}}E_{N}^{\ast}}{M_{N}^{\ast2}}
    - \ln\left(\frac{k_{F_{N}}+E_{N}^{\ast}}{M_{N}^{\ast}}\right) \right],
    \label{eq:scalar-density} \\
  \rho_{N}
  & = \braket{\psi_{N}^{\dagger}\psi_{N}}
    \nonumber \\
  & = \frac{1}{\pi^{2}} \int_{0}^{k_{F_{N}}} dk \, k^{2}
    = \frac{k_{F_{N}}^{3}}{3\pi^{2}},
    \label{eq:baryon-density}
\end{align}
where $k_{F_{N}}$ and $E_{N}^{\ast}\left(=\sqrt{k_{F_{N}}^{2}+M_{N}^{\ast2}}\right)$ are the Fermi momentum and energy for $N$.
With the self-consistent calculations of the meson fields, $\varepsilon_{B}$ and $P_{B}$ are respectively given by $\varepsilon_{B}=\sum_{N}\varepsilon_{N}+\varepsilon_{M}$ and $P_{B}=\sum_{N}P_{N}+P_{M}$ where the nucleon and meson parts are expressed as
\begin{align}
  \varepsilon_{N}
  & = \frac{1}{\pi^{2}} \int_{0}^{k_{F_{N}}} dk \, k^{2} \sqrt{k^{2}+M_{N}^{\ast2}}
    = \frac{1}{4}\left(3E_{N}^{\ast}\rho_{N}+M_{N}^{\ast}\rho_{N}^{s}\right),
    \label{eq:energy-density-nucleon} \\
  P_{N}
  & = \frac{1}{3\pi^{2}} \int_{0}^{k_{F_{N}}} dk \, \frac{k^{4}}{\sqrt{k^{2}+M_{N}^{\ast2}}}
    = \frac{1}{4}\left(E_{N}^{\ast}\rho_{N}-M_{N}^{\ast}\rho_{N}^{s}\right),
    \label{eq:pressure-nucleon}
\end{align}
and
\begin{align}
  \varepsilon_{M}
  & = \frac{1}{2}\left(m_{\sigma}^{2}\bar{\sigma}^{2}+m_{\omega}^{2}\bar{\omega}^{2}+m_{\delta}^{2}\bar{\delta}^{2}+m_{\rho}^{2}\bar{\rho}^{2}\right)
    \nonumber \\
  & + \frac{1}{3}g_{2}\bar{\sigma}^{3} + \frac{1}{4}g_{3}\bar{\sigma}^{4} + \frac{3}{4}c_{3}\bar{\omega}^{4}
    + \frac{1}{4}\bar{\delta}^{4} + \frac{3}{4}e_{3}\bar{\rho}^{4}
    - \Gamma_{\sigma\delta}\bar{\sigma}\bar{\delta}^{2} - \Lambda_{\sigma\delta}\bar{\sigma}^{2}\bar{\delta}^{2}
    + 3\Lambda_{\omega\rho}\bar{\omega}^{2}\bar{\rho}^{2},
    \label{eq:energy-density-meson} \\
  P_{M}
  & = -\frac{1}{2}\left(m_{\sigma}^{2}\bar{\sigma}^{2}-m_{\omega}^{2}\bar{\omega}^{2}+m_{\delta}^{2}\bar{\delta}^{2}-m_{\rho}^{2}\bar{\rho}^{2}\right)
    \nonumber \\
  & - \frac{1}{3}g_{2}\bar{\sigma}^{3} - \frac{1}{4}g_{3}\bar{\sigma}^{4} + \frac{1}{4}c_{3}\bar{\omega}^{4}
    - \frac{1}{4}\bar{\delta}^{4} + \frac{1}{4}e_{3}\bar{\rho}^{4}
    + \Gamma_{\sigma\delta}\bar{\sigma}\bar{\delta}^{2} + \Lambda_{\sigma\delta}\bar{\sigma}^{2}\bar{\delta}^{2}
    + \Lambda_{\omega\rho}\bar{\omega}^{2}\bar{\rho}^{2}.
    \label{eq:pressure-meson}
\end{align}

\subsection{Nuclear bulk properties} \label{sec:bulk}

In general, the bulk properties of infinite nuclear matter are identified by the expansion of isospin-asymmetric nuclear EoS with a power series in the isospin asymmetry, $\alpha=(\rho_{n}-\rho_{p})/\rho_{B}$, and the total baryon density, $\rho_{B}\left(=\rho_{n}+\rho_{p}\right)$~\citep{Chen:2007ih,Chen:2009wv}.
The binding energy per nucleon is then written as
\begin{align}
  E_{B}(\rho_{B},\alpha)
  & = \frac{\varepsilon_{B}(\rho_{B},\alpha)}{\rho_{B}} - M_{N}
    \nonumber \\
  & = E_{0}(\rho_{B}) + E_{\rm sym}(\rho_{B})\alpha^{2} + \mathcal{O}(\alpha^{4}),
    \label{eq:binding-energy}
\end{align}
where $E_{0}(\rho_{B})$ is the binding energy per nucleon of symmetric nuclear matter (SNM) and $E_{\rm sym}(\rho_{B})$ is the nuclear symmetry energy (NSE),
\begin{equation}
  E_{\rm sym}(\rho_{B})
  = \frac{1}{2} \left. \frac{\partial^{2}E_{B}(\rho_{B},\alpha)}{\partial\alpha^{2}} \right|_{\alpha=0}.
  \label{eq:definition-Esym}
\end{equation}
Besides, $E_{0}(\rho_{B})$ and $E_{\rm sym}(\rho_{B})$ can be expanded around the nuclear saturation density, $\rho_{0}$, as
\begin{align}
  E_{0}(\rho_{B})
  & = E_{0}(\rho_{0}) + \frac{K_{0}}{2}\chi^{2} + \frac{J_{0}}{6}\chi^{3} + \mathcal{O}(\chi^{4}),
    \label{eq:Binding-energy-sym} \\
  E_{\rm sym}(\rho_{B})
  & = E_{\rm sym}(\rho_{0}) + L\chi + \frac{K_{\rm sym}}{2}\chi^{2} + \frac{J_{\rm sym}}{6}\chi^{3} + \mathcal{O}(\chi^{4}),
    \label{eq:Binding-energy-asym}
\end{align}
with $\chi=(\rho_{B}-\rho_{0})/3\rho_{0}$ being the dimensionless variable characterizing the deviations of $\rho_{B}$ from $\rho_{0}$.
The incompressibility coefficient of SNM, $K_{0}$, the slope and curvature parameters of NSE, $L$ and $K_{\rm sym}$, and the third-order incompressibility coefficients of SNM and NSE, $J_{0}$ and $J_{\rm sym}$, are respectively defined as
\begin{align}
  K_{0} = 9 \rho_{B}^{2}
  & \left. \frac{d^{2}E_{0}(\rho_{B})}{d\rho_{B}^{2}} \right|_{\rho_{B}=\rho_{0}},
    \label{eq:definition-K0}\\
  L = 3 \rho_{B} \left. \frac{d E_{\rm sym}(\rho_{B})}{d\rho_{B}} \right|_{\rho_{B}=\rho_{0}},
  & \quad K_{\rm sym} = 9 \rho_{B}^{2} \left. \frac{d^{2}E_{\rm sym}(\rho_{B})}{d\rho_{B}^{2}} \right|_{\rho_{B}=\rho_{0}},
    \label{eq:definition-L-Ksym} \\
  J_{0} = 27 \rho_{B}^{3} \left. \frac{d^{3}E_{0}(\rho_{B})}{d\rho_{B}^{3}} \right|_{\rho_{B}=\rho_{0}},
  & \quad  J_{\rm sym} = 27 \rho_{B}^{3} \left. \frac{d^{3}E_{\rm sym}(\rho_{B})}{d\rho_{B}^{3}} \right|_{\rho_{B}=\rho_{0}}.
\end{align}

Taking into account the thermodynamic condition, the pressure of infinite nuclear matter, $P_{B}(\rho_{B},\alpha)$, is given by
\begin{align}
  P_{B}(\rho_{B},\alpha)
  & = \rho_{B}^{2} \frac{\partial E_{B}(\rho_{B},\alpha)}{\partial\rho_{B}}
    \nonumber \\
  & = \rho_{B}^{2} \frac{\partial}{\partial\rho_{B}}\left[ \frac{\varepsilon_{B}(\rho_{B},\alpha)}{\rho_{B}}-M_{N} \right]
    \nonumber \\
  & = \rho_{B} \frac{\partial\varepsilon_{B}(\rho_{B},\alpha)}{\partial\rho_{B}} - \varepsilon_{B}(\rho_{B},\alpha),
    \label{eq:definition-pressure}
\end{align}
with the binding energy per nucleon in Eq.~\eqref{eq:binding-energy}.
The nuclear incompressibility, $K_{B}(\rho_{B},\alpha)$, is then expressed as
\begin{align}
  K_{B}(\rho_{B},\alpha)
  & = 9 \rho_{B}^{2} \frac{\partial^{2}E_{B}(\rho_{B},\alpha)}{\partial\rho_{B}^{2}}
    \nonumber \\
  & = 9 \rho_{B}^{2} \frac{\partial}{\partial\rho_{B}} \left[ \frac{P_{B}(\rho_{B},\alpha)}{\rho_{B}^{2}} \right]
    \nonumber \\
  & = 9 \left[ \frac{\partial P_{B}(\rho_{B},\alpha)}{\partial\rho_{B}} - 2 \frac{P_{B}(\rho_{B},\alpha)}{\rho_{B}} \right].
    \label{eq:definition-incompressibility}
\end{align}
Hence, the incompressibility coefficient of SNM, $K_{0}$, in Eq.~\eqref{eq:definition-K0} is related with $K_{B}$ through $K_{0}=K_{B}(\rho_{0},0)$.
In the RMF calculation, we can obtain the analytical expression of $K_{B}(\rho_{B},\alpha)$ using the following equation:
\begin{align}
  \frac{\partial P_{B}}{\partial\rho_{B}}
  & = \frac{1}{3\rho_{B}}\sum_{N=p,n}\rho_{N}\frac{k_{F_{N}}^{2}}{E_{N}^{\ast}}
    \nonumber \\
  & \quad
    - \sum_{N=p,n} \rho_{N} \frac{M_{N}^{\ast}}{E_{N}^{\ast}} \left[ g_{\sigma}\frac{\partial\bar{\sigma}}{\partial\rho_{B}}
    + g_{\delta} (\bm{\tau}_{N})_{3} \frac{\partial\bar{\delta}}{\partial\rho_{B}} \right]
    + m_{\omega}^{\ast2}\bar{\omega} \frac{\partial\bar{\omega}}{\partial\rho_{B}}
    + m_{\rho}^{\ast2}\bar{\rho} \frac{\partial\bar{\rho}}{\partial\rho_{B}},
\end{align}
where the density derivatives of meson fields are calculated through the relation
\begin{equation}
  \frac{\partial\mathcal{M}}{\partial\rho_{B}}
  = \sum_{N=p,n} \frac{\rho_{N}}{\rho_{B}} \frac{\partial\mathcal{M}}{\partial\rho_{N}}
  \quad
  \left( \mathcal{M} = \bar{\sigma}, \bar{\omega}, \bar{\delta}, \bar{\rho} \right),
\end{equation}
with
\begin{align}
  \frac{\partial\bar{\sigma}}{\partial\rho_{N}}
  = \frac{M_{N}^{\ast}}{E_{N}^{\ast}} \frac{G_{\sigma}+G_{\delta}(\bm{\tau}_{N})_{3}H_{\sigma\delta}}{1-H_{\sigma\delta}H_{\delta\sigma}},
  & \qquad
    \frac{\partial\bar{\omega}}{\partial\rho_{N}}
    = \frac{G_{\omega}-G_{\rho}(\bm{\tau}_{N})_{3}H_{\omega\rho}}{1-H_{\omega\rho}H_{\rho\omega}},
    \label{eq:derivative1} \\
  \frac{\partial\bar{\delta}}{\partial\rho_{N}}
  = \frac{M_{N}^{\ast}}{E_{N}^{\ast}} \frac{G_{\sigma}H_{\delta\sigma}+G_{\delta}(\bm{\tau}_{N})_{3}}{1-H_{\sigma\delta}H_{\delta\sigma}},
  & \qquad
    \frac{\partial\bar{\rho}}{\partial\rho_{N}}
    = \frac{-G_{\omega}H_{\rho\omega}+G_{\rho}(\bm{\tau}_{N})_{3}}{1-H_{\omega\rho}H_{\rho\omega}},
    \label{eq:derivative2}
\end{align}
and
\begin{equation}
  (\bm{\tau}_{N})_{3}= \left\{
    \begin{array}{c}
      +1 \\
      -1 \\
    \end{array} \right.
  \ \text{for} \ N = {p \choose n}.
\end{equation}
We here use the following quantities:
% \begin{align}
%   G_{\sigma} = \frac{g_{\sigma}}{M_{\sigma}^{2}}, \quad
%   G_{\omega} = \frac{g_{\omega}}{M_{\omega}^{2}}, \quad
%   G_{\delta} = \frac{g_{\delta}}{M_{\delta}^{2}}, \quad
%   G_{\rho}   = \frac{g_{\rho}  }{M_{\rho}^{2}  }, \quad \qquad \\
%   H_{\sigma\delta} = \frac{L_{\sigma\delta}^{2}}{M_{\sigma}^{2}}, \quad
%   H_{\delta\sigma} = \frac{L_{\sigma\delta}^{2}}{M_{\delta}^{2}}, \quad
%   H_{\omega\rho} = \frac{4\Lambda_{\omega\rho}\bar{\omega}\bar{\rho}}{M_{\omega}^{2}}, \quad
%   H_{\rho\omega} = \frac{4\Lambda_{\omega\rho}\bar{\omega}\bar{\rho}}{M_{\rho}^{2}},
% \end{align}
\begin{equation}
  G_{\sigma} = \frac{g_{\sigma}}{M_{\sigma}^{2}}, \quad
  G_{\omega} = \frac{g_{\omega}}{M_{\omega}^{2}}, \quad
  G_{\delta} = \frac{g_{\delta}}{M_{\delta}^{2}}, \quad
  G_{\rho}   = \frac{g_{\rho}  }{M_{\rho}^{2}  },
\end{equation}
and
\begin{equation}
  H_{\sigma\delta} = \frac{L_{\sigma\delta}^{2}}{M_{\sigma}^{2}}, \quad
  H_{\delta\sigma} = \frac{L_{\sigma\delta}^{2}}{M_{\delta}^{2}}, \quad
  H_{\omega\rho} = \frac{4\Lambda_{\omega\rho}\bar{\omega}\bar{\rho}}{M_{\omega}^{2}}, \quad
  H_{\rho\omega} = \frac{4\Lambda_{\omega\rho}\bar{\omega}\bar{\rho}}{M_{\rho}^{2}},
\end{equation}
with
\begin{align}
  M_{\sigma}^{2}(\bar{\sigma},\bar{\delta})
  & = m_{\sigma}^{\ast2}(\bar{\sigma},\bar{\delta})
    + g_{2}\bar{\sigma} + 2g_{3}\bar{\sigma}^{2} + \Gamma_{\sigma\delta}\bar{\delta}^{2}/\bar{\sigma} + g_{\sigma}^{2}\left(J_{p}+J_{n}\right),\\
  M_{\omega}^{2}(\bar{\omega},\bar{\rho})
  & = m_{\omega}^{\ast2}(\bar{\omega},\bar{\rho}) + 2c_{3}\bar{\omega}^{2}, \\
  M_{\delta}^{2}(\bar{\sigma},\bar{\delta})
  & = m_{\delta}^{\ast2}(\bar{\sigma},\bar{\delta}) + 2d_{3}\bar{\delta}^{2} + g_{\delta}^{2}\left(J_{p}+J_{n}\right),\\
  M_{\rho}^{2}(\bar{\omega},\bar{\rho})
  & = m_{\rho}^{\ast2}(\bar{\omega},\bar{\rho}) + 2e_{3}\bar{\rho}^{2}, \\
  L_{\sigma\delta}^{2}(\bar{\sigma},\bar{\delta})
  & = 2\Gamma_{\sigma\delta}\bar{\delta} + 4\Lambda_{\sigma\delta}\bar{\sigma}\bar{\delta} - g_{\sigma}g_{\delta}\left(J_{p}-J_{n}\right),
\end{align}
where the effective meson masses, $m_{\sigma}^{\ast2}$, $m_{\omega}^{\ast2}$, $m_{\delta}^{\ast2}$ and $m_{\rho}^{\ast2}$, are given in Eqs.~\eqref{eq:sigma-mass}--\eqref{eq:rho-mass}, and $J_{N}$ for $N~\left(=p,n\right)$ reads
\begin{equation}
  J_{N} = 3\left( \frac{\rho_{N}^{s}}{M_{N}^{\ast}} - \frac{\rho_{N}}{E_{N}^{\ast}} \right).
\end{equation}

According to the Hugenholtz-Van Hove theorem in nuclear matter, $E_{\rm sym}$ defined in Eq.~\eqref{eq:definition-Esym} can be generally written as
\begin{align}
  E_{\rm sym}(\rho_{B})
  & = \frac{1}{2} \frac{\partial}{\partial\alpha} \left[ \frac{\partial E_{B}(\rho_{B},\alpha)}{\partial\alpha} \right]_{\alpha=0}
    \nonumber \\
  & = \frac{1}{8}\rho_{B} \left. \left(\frac{\partial}{\partial\rho_{p}}-\frac{\partial}{\partial\rho_{n}}\right)
      \left[E_{p}(k_{F_{p}})-E_{n}(k_{F_{n}})\right] \right|_{\rho_{p}=\rho_{n}},
\end{align}
where $E_{N}$ is the single-particle energy for $N$, which is determined self-consistently by solving the following transcendental equation~\citep{Czerski:2002pz,Cai:2012en}:
\begin{equation}
  E_{N}(k) = \left[E_{N}^{\ast}(k)-\Sigma_{N}^{0}(k)\right]_{k^{0}=E_{N}(k)}.
\end{equation}
The effective mass, (four) momentum, and energy for $N$ are here defined as~\citep{Miyatsu:2011bc,Katayama:2012ge}
\begin{align}
  M_{N}^{\ast}(k)
  & = M_{N} + \Sigma_{N}^{s}(k), \\
  k_{N}^{\mu\ast}
  & = (k_{N}^{\ast0},\bm{k}_{N}^{\ast})
    \nonumber \\
  & = (k^{0}+\Sigma_{N}^{0}(k),\bm{k}+\hat{k}\Sigma_{N}^{v}(k)), \\
  E_{N}^{\ast}(k)
  & = \sqrt{\bm{k}_{N}^{\ast2}+M_{N}^{\ast2}(k)},
\end{align}
with $\Sigma_{N}^{s(0)[v]}$ being the scalar (time) [space] component of nucleon self-energy.
In addition, $E_{\rm sym}$ is divided into the kinetic and potential terms as
\begin{equation}
  E_{\rm sym}(\rho_{B})=E_{\rm sym}^{\rm kin}(\rho_{B})+E_{\rm sym}^{\rm pot}(\rho_{B}).
\end{equation}
Based on the Lorentz-covariant decomposition of NSE~\citep{Miyatsu:2020vzi}, $E_{\rm sym}^{\rm pot}$ is expressed as
\begin{equation}
  E_{\rm sym}^{\rm pot}(\rho_{B})=E_{\rm sym}^{s}(\rho_{B})+E_{\rm sym}^{0}(\rho_{B})+E_{\rm sym}^{v}(\rho_{B}),
\end{equation}
with the scalar ($s$), time ($0$), and space ($v$) components.
The $E_{\rm sym}$ is thus computed as follows:
\begin{align}
  E_{\rm sym}^{\rm kin}(\rho_{B})
  & = \frac{1}{6}\frac{k_{F}^{\ast}}{E_{F}^{\ast}}k_{F},
  \label{eq:Esym-kin} \\
  E_{\rm sym}^{s}(\rho_{B})
  & = \frac{1}{8}\rho_{B} \frac{M_{F}^{\ast}}{E_{F}^{\ast}} \left. \left(\frac{\partial}{\partial\rho_{p}}-\frac{\partial}{\partial\rho_{n}}\right)
    \left(\Sigma_{p}^{s}-\Sigma_{n}^{s}\right) \right|_{\rho_{p}=\rho_{n}},
    \label{eq:Esym-scalar} \\
  E_{\rm sym}^{0}(\rho_{B})
  & = -\frac{1}{8}\rho_{B} \left. \left(\frac{\partial}{\partial\rho_{p}}-\frac{\partial}{\partial\rho_{n}}\right)
    \left(\Sigma_{p}^{0}-\Sigma_{n}^{0}\right) \right|_{\rho_{p}=\rho_{n}},
    \label{eq:Esym-time} \\
  E_{\rm sym}^{v}(\rho_{B})
  & = \frac{1}{8}\rho_{B} \frac{k_{F}^{\ast}}{E_{F}^{\ast}} \left. \left(\frac{\partial}{\partial\rho_{p}}-\frac{\partial}{\partial\rho_{n}}\right)
    \left(\Sigma_{p}^{v}-\Sigma_{n}^{v}\right) \right|_{\rho_{p}=\rho_{n}},
    \label{eq:Esym-space}
\end{align}
where the effective quantities at the Fermi surface in Eq.~\eqref{eq:Esym-kin}--\eqref{eq:Esym-space} are then given by $M_{F}^{\ast}=M_{p}^{\ast}(k_{F})=M_{n}^{\ast}(k_{F})$, $k_{F}^{\ast}=|\bm{k}_{p}^{\ast}(k_{F})|=|\bm{k}_{n}^{\ast}(k_{F})|$, and $E_{F}^{\ast}=E_{p}^{\ast}(k_{F})=E_{n}^{\ast}(k_{F})$ at $\rho_{p}=\rho_{n}$, namely $k_{F_{p}}=k_{F_{n}}=k_{F}$.
In RMF approximation, $\Sigma_{N}^{s,0,v}$ are respectively given by
\begin{align}
  \Sigma_{N}^{s}
  & = -g_{\sigma}\bar{\sigma} - g_{\delta}(\bm{\tau}_{N})_{3}\bar{\delta}, \\
  \Sigma_{N}^{0}
  & = -g_{\omega}\bar{\omega} - g_{\rho}(\bm{\tau}_{N})_{3}\bar{\rho}, \\
  \Sigma_{N}^{v}
  & = 0.
\end{align}
Using Eqs.~\eqref{eq:derivative1} and \eqref{eq:derivative2}, $E_{\rm sym}$ can be finally expressed as
\begin{align}
  E_{\rm sym}(\rho_{B})
  % & = E_{\rm sym}^{\rm kin}(\rho_{B}) + E_{\rm sym}^{\rm pot}(\rho_{B})
  %   \nonumber \\
  & = E_{\rm sym}^{\rm kin}(\rho_{B}) + E_{\rm sym}^{s}(\rho_{B}) + E_{\rm sym}^{0}(\rho_{B})
    \nonumber \\
  & = \frac{1}{6}\frac{k_{F}^{2}}{E_{F}^{\ast}}
    - \frac{1}{2}\frac{g_{\delta}^{2}}{M_{\delta}^{2}(\bar{\sigma},0)} \left(\frac{M_{F}^{\ast}}{E_{F}^{\ast}}\right)^{2}\rho_{B}
    + \frac{1}{2}\frac{g_{\rho}^{2}}{M_{\rho}^{2}(\bar{\omega},0)}\rho_{B}.
    \label{eq:Esym-RMF}
\end{align}
Note that $k_{F}^{\ast}=k_{F}$ and $E_{\rm sym}^{v}(\rho_{B})=0$ in RMF approximation.

The $L$ and $K_{\rm sym}$ given in Eq.~\eqref{eq:definition-L-Ksym} are also expressed as
\begin{align}
  L
  & = L^{\rm kin} + L^{\rm pot}
    \nonumber \\
  & = L^{\rm kin}+L^{s}+L^{0}, \\
  K_{\rm sym}
  & = K_{\rm sym}^{\rm kin} + K_{\rm sym}^{\rm pot}
    \nonumber \\
  & = K_{\rm sym}^{\rm kin}+K_{\rm sym}^{s}+K_{\rm sym}^{0},
\end{align}
where the kinetic, scalar, and time components are respectively given by
\begin{align}
  L^{\rm kin}
  & = E_{\rm sym}^{\rm kin}(\rho_{0}) \left[ 1+\left(\frac{M_{F}^{\ast}}{E_{F}^{\ast}}\right)^{2}\mathcal{K}_{B}(\rho_{0}) \right], \\
  L^{s(0)}
  & = 3E_{\rm sym}^{s(0)}(\rho_{0}) \left[ 1-\rho_{0}\,\mathcal{T}_{B}^{s(0)}(\rho_{0}) \right], \\
% \end{align}
% and
% \begin{align}
  K_{\rm sym}^{\rm kin}
  & = -2L^{\rm kin} + \left(\frac{M_{F}^{\ast}}{E_{F}^{\ast}}\right)^{2} \left[ E_{\rm sym}^{\rm kin}(\rho_{0})\mathcal{N}_{B}(\rho_{0})+L^{\rm kin}\mathcal{K}_{B}(\rho_{0}) \right], \\
  K_{\rm sym}^{\rm s(0)}
  & = 3L^{s(0)} \left[1-\rho_{0}\,\mathcal{T}_{B}^{s(0)}(\rho_{0})\right]
    - 9E_{\rm sym}^{s(0)} \left[ 1+\rho_{0}^{2} \left.\frac{d\,\mathcal{T}_{B}^{s(0)}(\rho_{B})}{d\rho_{B}}\right|_{\rho_{B}=\rho_{0}} \right],
\end{align}
with
\begin{align}
  \mathcal{K}_{B}(\rho_{B})
  & = 1 + 3\frac{g_{\sigma}^{2}}{M_{\sigma}^{2}(\bar{\sigma},0)}\frac{\rho_{B}}{E_{F}^{\ast}}, \\
  \mathcal{T}_{B}^{s}(\rho_{B})
  & = \frac{2}{3\rho_{B}}\left(\frac{k_{F}}{E_{F}^{\ast}}\right)^{2}\mathcal{K}_{B}(\rho_{B})
    - \frac{g_{\sigma}\left(2\Gamma_{\sigma\delta}+4\Lambda_{\sigma\delta}\bar{\sigma}\right)}
      {M_{\sigma}^{2}(\bar{\sigma},0)M_{\delta}^{2}(\bar{\sigma},0)}
      \frac{M_{F}^{\ast}}{E_{F}^{\ast}} \nonumber \\
  & \quad
    + \frac{g_{\delta}^{2}}{M_{\delta}^{2}(\bar{\sigma},0)}
      \left[ \left(\frac{k_{F}}{E_{F}^{\ast}}\right)^{2}\frac{\mathcal{K}_{B}(\rho_{B})}{E_{F}^{\ast}}
    - 2\frac{g_{\sigma}^{2}}{M_{\sigma}^{2}(\bar{\sigma},0)}\frac{J_{p}+J_{n}}{E_{F}^{\ast}} \right], \\
  \mathcal{T}_{B}^{0}(\rho_{B})
  & = \frac{4g_{\omega}\Lambda_{\omega\rho}\bar{\omega}}{M_{\omega}^{2}(\bar{\omega},0)M_{\rho}^{2}(\bar{\omega},0)}, \\
  \mathcal{N}_{B}(\rho_{B})
  & = 3\rho_{B}\frac{d\mathcal{K}_{B}(\rho_{B})}{d\rho_{B}} - 2\left(\frac{k_{F}}{E_{F}^{\ast}}\right)^{2}\mathcal{K}_{B}^{2}(\rho_{B}).
\end{align}

\subsection{Stability of nuclear and neutron star matter} \label{sec:stability}

In order to move on the calculations of neutron stars in which the charge neutrality and $\beta$ equilibrium conditions are imposed, we introduce the degrees of freedom of leptons (electrons and muons) as well as nucleons and mesons in Eq.~\eqref{eq:Ltot}
\begin{equation}
  \mathcal{L}_{L} = \bar{\psi}_{\ell}\left(i\gamma_{\mu}\partial^{\mu}-\hat{m}_{\ell}\right)\psi_{\ell},
\end{equation}
where $\psi_{\ell}={\psi_{e} \choose \psi_{\mu}}$ is the lepton field and its mass is given by $\hat{m}_{\ell}=\mathrm{diag}(m_{e},m_{\mu})$.

When we consider the stability of matter in cold neutron stars, the first principle of thermodynamics should be considered:
\begin{equation}
  du = -Pdv-\mu dq,
\end{equation}
with $u$, $P$, $v \left(=1/\rho_{B}\right)$, $\mu$, and $q$ being the total internal energy per nucleon, pressure, volume per nucleon, chemical potential, and charge fraction, respectively~\citep{Kubis:2006kb,Lattimer:2006xb,Xu:2009vi,Moustakidis:2010zx}.
In neutron star matter, the charge neutrality and $\beta$ equilibrium conditions read
\begin{align}
  \mu
  & = \mu_{n}-\mu_{p} = \mu_{e} = \mu_{\mu}, \\
  q
  & = Y_{p} - Y_{L} = \rho_{p}/\rho_{B} - \sum_{\ell=e,\mu}\rho_{\ell}/\rho_{B} = 0,
\end{align}
with $\rho_{\ell}$ the lepton density.
The stability of neutron star matter are then expressed as the following two constraints on chemical potential and pressure:
\begin{align}
  -\left(\frac{\partial\mu}{\partial q}\right)_{v}
  & > 0,
    \label{eq:constraint-chemical-potential} \\
  -\left(\frac{\partial P}{\partial v}\right)_{\mu}
  & > 0.
    \label{eq:constraint-pressure}
\end{align}

The total internal energy per baryon, $u(v,q)$, can be decomposed into the baryon ($B$) and lepton ($L$) contributions as
\begin{equation}
  u(v,q)
  = u(\rho_{B},\alpha)
  = E_{B}(\rho_{B},\alpha) + E_{L}(\rho_{B},\alpha),
\end{equation}
with $\alpha=1-2Y_{p}$.
At zero temperature, the $\beta$ equilibrium condition leads to the relation~\citep{Psonis:2006kz}
\begin{align}
  \mu
  & = -\left(\frac{\partial E_{B}}{\partial Y_{p}}\right)_{\rho_{B}}
    \nonumber \\
  & = 2\left(\frac{\partial E_{B}}{\partial\alpha}\right)_{\rho_{B}}
    = 2E_{\rm ISB}(\rho_{B},\alpha). %\Bigr|_{\rho_{B}\text{: fixed}},
\end{align}
where the isospin symmetry breaking (ISB) energy of infinite nuclear matter is given by
\begin{align}
  E_{\rm ISB}(\rho_{B},\alpha)
  & = \left[ \frac{\partial E_{B}(\rho_{B},\alpha)}{\partial\alpha} \right]_{\rho_{B}}
    \nonumber \\
  & = \frac{1}{2}\left[E_{n}(k_{F_{n}})-E_{p}(k_{F_{p}})\right].
  % = -\frac{1}{2}\sum_{N=p,n}(\bm{\tau}_{N})_{3}\left(E_{N}^{\ast}-\Sigma_{N}^{0}\right).
\end{align}
Considering the differentiation of $\mu(v,q) \left(=\mu(\rho_{B},\alpha)\right)$, we find
\begin{align}
  -\left(\frac{\partial q}{\partial\mu}\right)_{v}
  & = \frac{1}{2}\left(\frac{\partial\alpha}{\partial\mu}\right)_{\rho_{B}}
    + \frac{1}{\rho_{B}}\sum_{\ell=e,\mu}\left(\frac{\partial\rho_{\ell}}{\partial\mu}\right)_{\rho_{B}}
    \nonumber \\
  & = \frac{1}{8E_{\rm sym}(\rho_{B},\alpha)} + \frac{\mu}{\pi^{2}\rho_{B}}(k_{F_{e}}+k_{F_{\mu}})
    =: V_{\mu}(\rho_{B},\alpha),
    \label{eq:chemical-potential-condition}
\end{align}
where $k_{F_{e}}$ and $k_{F_{\mu}}$ are respectively the Fermi momenta for electrons ($e$) and muons ($\mu$).
For simplicity, we here define the nuclear symmetry energy involving the isospin asymmetry, $\alpha$, as
\begin{align}
  E_{\rm sym}(\rho_{B},\alpha)
  & = \frac{1}{2}\frac{\partial^{2}E_{B}(\rho_{B},\alpha)}{\partial\alpha^{2}}
    \nonumber \\
  & = \frac{1}{2}\frac{\partial E_{\rm ISB}(\rho_{B},\alpha)}{\partial\alpha}.
\end{align}
Note that we explicitly keep $\alpha$ to consider the stability of nuclear and neutron star matter, though the nuclear symmetry energy is in general calculated at $\rho_{p}=\rho_{n}$, namely $\alpha=0$, as shown in Eq.~\eqref{eq:definition-Esym}.
Hence the stability constraint on chemical potential, $V_{\mu}(\rho_{B},\alpha)>0$, can be satisfied by assuming that $E_{\rm sym}(\rho_{B},\alpha)$ is positive at any $\rho_{B}$.

As for the pressure stability, the differentiation of $P(v,q) \left(=P(\rho_{B},\alpha)\right)$ reads
\begin{equation}
  -\left(\frac{\partial P}{\partial v}\right)_{\mu}
  = \rho_{B}^{2} \left[ \left(\frac{\partial P_{B}}{\partial\rho_{B}}\right)_{\mu}
    + \left(\frac{\partial P_{L}}{\partial\rho_{B}}\right)_{\mu} \right],
\end{equation}
with the baryon and lepton contributions.
Similar to Eq.~\eqref{eq:chemical-potential-condition}, the baryon contribution is given by
\begin{equation}
  \left(\frac{\partial P_{B}}{\partial\rho_{B}}\right)_{\mu}
  = 2\rho_{B}\frac{\partial E_{B}(\rho_{B},\alpha)}{\partial\rho_{B}}
  + \rho_{B}^{2}\frac{\partial^{2}E_{B}(\rho_{B},\alpha)}{\partial^{2}\rho_{B}}
  - \rho_{B}^{2} \left. \left[\frac{\partial^{2}E_{B}(\rho_{B},\alpha)}{\partial\rho_{B}\partial\alpha}\right]^{2} \right/ \frac{\partial^{2}E_{B}(\rho_{B},\alpha)}{\partial^{2}\alpha}.
  \label{eq:chemical-potential-condition-baryon}
\end{equation}
Using the thermodynamic definitions of pressure and incompressibility of infinite nuclear matter in Eqs.~\eqref{eq:definition-pressure} and \eqref{eq:definition-incompressibility}, this equation can be simplified as
\begin{equation}
  \left(\frac{\partial P_{B}}{\partial\rho_{B}}\right)_{\mu}
  = 2\frac{P_{B}(\rho_{B},\alpha)}{\rho_{B}}
  + \frac{1}{9}K_{B}(\rho_{B},\alpha)
  - \frac{1}{18}\frac{L_{\rm ISB}^{2}(\rho_{B},\alpha)}{E_{\rm sym}(\rho_{B},\alpha)},
  \label{eq:thermodynamic-condition}
\end{equation}
where the slope of ISB energy, $L_{\rm ISB}(\rho_{B},\alpha)$, is defined as
\begin{equation}
  L_{\rm ISB}(\rho_{B},\alpha)
  = 3\rho_{B}\frac{\partial E_{\rm ISB}(\rho_{B},\alpha)}{\partial\rho_{B}}.
\end{equation}
The lepton contribution is also given by the simple form under the $\beta$ equilibrium condition:
\begin{equation}
  \left(\frac{\partial P_{L}}{\partial\rho_{B}}\right)_{\mu}
  = \frac{\rho_{e}k_{F_{e}}^{2}+\rho_{\mu}k_{F_{\mu}}^{2}}{3\mu\rho_{B}}.
\end{equation}

Therefore, the stability of neutron star matter under the charge neutrality and $\beta$ equilibrium conditions can be clarified by the thermodynamic constraints on chemical potential and pressure, namely $V_{\mu}(\rho_{B},\alpha)>0$ and
\begin{equation}
  V_{P}(\rho_{B},\alpha) := \left(\frac{\partial P_{B}}{\partial\rho_{B}}\right)_{\mu}+\left(\frac{\partial P_{L}}{\partial\rho_{B}}\right)_{\mu}>0.
  \label{eq:pressure-condition}
\end{equation}
The thermodynamic stability is used in several calculations of nuclear and neutron star matter, for instance, the compressibility of $\beta$-equilibrated matter~\citep{Zabari:2018tjk,Kubis:2020ysv} and the phase transition between the crust and core regions in neutron stars~\citep{Routray:2016yvp,Zhang:2018vrx,Xie:2019sqb}.

% \clearpage

\section{Results and discussions} \label{sec:results}

\subsection{Nuclear models}

% \begin{landscape}
%   \input{./Tables/Table-CCs.tex}
%   \input{./Tables/Table-matter.tex}
% \end{landscape}
\input{./Tables/Table-CCs.tex}
\input{./Tables/Table-matter.tex}
\input{./Tables/Table-CCs-delta.tex}
\input{./Tables/Table-CCs-rho.tex}

We adopt the recently developed effective interactions labeled as the OMEG family, which are constructed to reproduce the characteristics of finite nuclei, nuclear matter, and neutron stars~\citep{Miyatsu:2022bhh,Miyatsu:2023lki}.
In particular, the $\delta$--$N$ coupling and $\sigma$--$\delta$ mixing in the OMEG family are determined so as to support the astrophysical constraints on the neutron-star radii from the NICER mission~\citep{Miller:2019cac,Miller:2021qha,Riley:2019yda,Riley:2021pdl} and the tidal deformabilities from the binary merger events due to GW signals~\citep{LIGOScientific:2018cki,LIGOScientific:2020zkf}.
Various theoretical calculations using the well-calibrated parameter sets based on the RMF models are also presented: BigApple~\citep{Fattoyev:2020cws}, DINO~\citep{Reed:2023cap}, FSU-$\delta$~\citep{Li:2022okx}, FSUGarnet~\citep{Chen:2014mza}, FSUGold~\citep{Fattoyev:2010tb}, FSUGold2~\citep{Chen:2014sca}, Bayesian refinement of FSUGarnet and FSUGold2, FSUGarnet+R and FSUGold2+R~\citep{Salinas:2023nci,Salinas:2023epr}, HPNL0 and HPNL5~\citep{PhysRevC.108.055802}, IOPB-I~\citep{Kumar:2017wqp}, IU-FSU~\citep{Fattoyev:2010mx}, NL3~\citep{Lalazissis:1996rd}, PD15~\citep{Liliani:2021jne}, TAMUC-FSUa~\citep{Fattoyev:2013yaa,Piekarewicz:2013bea}, and TM1~\citep{Sugahara:1993wz}.
In Tables~\ref{tab:CCs} and \ref{tab:matter}, we summarize the model parameters and the properties of symmetric nuclear matter at $\rho_{0}$ for the effective interactions used in the present study.

In addition, we present the extended interactions based on the FSUGarnet, TAMUC-FSUa, and FSUGold2 models, in which the $\delta$--$N$ coupling are introduced to investigate the effect of $\delta$ meson.
% The $\delta$--$N$ coupling constant, $g_{\delta}$, is here fixed so as to maintain the result of $L$ from the original models, while the other parameters are readjusted to reproduce as much as possible the other characteristics, such as the binding energy per nucleon and charge radius of several closed-shell nuclei, and the saturation properties of nuclear matter.
\RevB{Since the $\delta$--$N$ coupling only influences the properties of $N\not=Z$ finite nuclei and isospin-asymmetric nuclear matter, we adjust $g_{\rho}^{2}$ and $\Lambda_{\omega\rho}$ to preserve the original model's predictions for $L$ when the $\delta$--$N$ coupling is included. Simultaneously, the other coupling constants related to the properties of $N=Z$ finite nuclei and isospin-symmetric nuclear matter---$g_{\sigma}^{2}$, $g_{\omega}^{2}$, $g_{2}$, and $g_{3}$---are readjusted to closely match the experimental data for the binding energy per nucleon and charge radius of several closed-shell nuclei, as well as to maintain the original $K_{0}$ value.}
The resultant coupling constants and nuclear properties for the FSUGarnet, TAMUC-FSUa, and FSUGold2 series are listed in Table~\ref{tab:CCs2}.
Furthermore, the parameter sets for the FSUGold2 with the $\delta$--$N$ coupling and the quartic self-interaction of $\rho$ meson are also given in Table~\ref{tab:CCs3}, where the quartic coupling constant, $e_{3}$, is varied in the range of $0\le e_{3}\le800$ with the fixed parameters, $c_{3}=144.12$ and $g_{\delta}^{2}=300$.

% \clearpage

\subsection{Finite nuclei}

\begin{figure}[b!]
  \begin{center}
    \includegraphics[width=8.8cm]{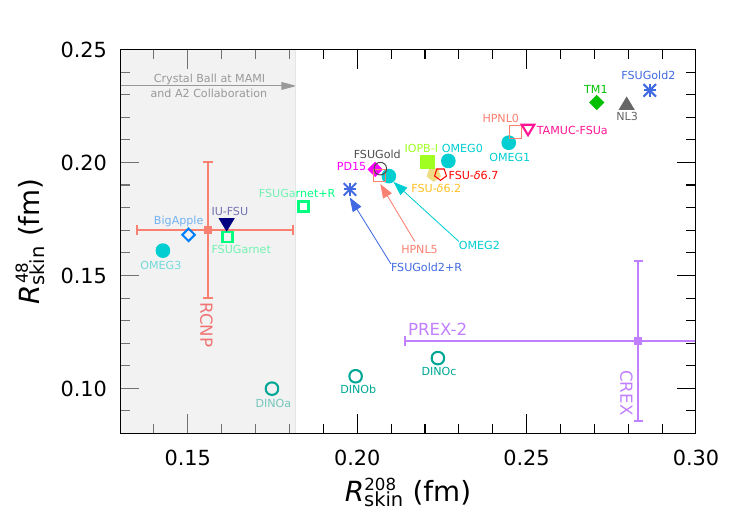}
    \includegraphics[width=8.8cm]{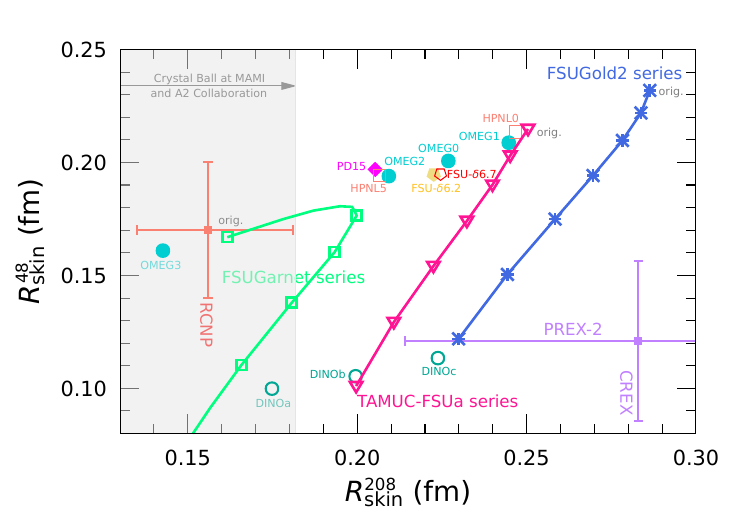}
  \end{center}
  \caption{Neutron skin thickness of $^{40}$Ca and $^{208}$Pb, $R_{\rm skin}^{48}$ and $R_{\rm skin}^{208}$.
    The left panel shows the results from the effective interactions presented in Tables~\ref{tab:CCs} and \ref{tab:matter}.
    The right panel is for the FSUGarnet, TAMUC-FSUa, and FSUGold2 series in Table~\ref{tab:CCs2}.}
  \label{fig:NST}
\end{figure}
The theoretical predictions for the neutron skin thickness of $^{40}$Ca and $^{208}$Pb, $R_{\rm skin}^{48}$ and $R_{\rm skin}^{208}$, in the RMF models are presented in Fig.~\ref{fig:NST}, compared with the experimental data: the electric dipole polarizability of $^{48}$Ca (RCNP; $R_{\rm skin}^{48}=0.14$--$0.20$~fm)~\citep{Birkhan:2016qkr}, the complete electric dipole response on $^{208}$Pb (RCNP; $R_{\rm skin}^{208}=0.156_{-0.021}^{+0.025}$~fm)~\citep{Tamii:2011pv}, the coherent pion photoproduction cross sections measurement of $^{208}$Pb (MAMI; $R_{\rm skin}^{208}=0.15\pm0.03(\mathrm{stat.})_{-0.03}^{+0.01}(\mathrm{sys.})$~fm)~\citep{Tarbert:2013jze}, and the parity-violating electron scattering off $^{48}$Ca (CREX; $R_{\rm skin}^{48}=0.121\pm0.026(\mathrm{exp.})\pm0.024(\mathrm{model})$~fm)~\citep{CREX:2022kgg} and off $^{208}$Pb (PREX-2; $R_{\rm skin}^{208}=0.283\pm0.071$~fm)~\citep{PREX:2021umo}.

\input{./Tables/Table-NST-all.tex}
As for the OMEG family, the OMEG0 and OMEG1 give the large values, $R_{\rm skin}^{208}=0.227$~fm and $R_{\rm skin}^{208}=0.245$~fm respectively, which meet the PREX-2 result.
The OMEG2 is selected so as to match the predicted result, $R_{\rm skin}^{208}=0.19\pm0.02$~fm, by the assessment of the theoretical uncertainty on parity-violating asymmetry in $^{208}$Pb~\citep{Reinhard:2021utv}.
Meanwhile, the OMEG3 exhibits the small value, $R_{\rm skin}^{48}=0.161$~fm, which satisfies the experimental result in RCNP and is near the range of CREX experiment, $R_{\rm skin}^{48}=0.121\pm0.035$~fm.
We summarize the predictions for the charge radius, $R_{\rm ch}$, neutron skin thickness, $R_{\rm skin}$, and weak radius, $R_{\rm wk}$, of $^{48}$Ca and $^{208}$Pb in Table~\ref{tab:NST}.
We here consider the zero-point energy correction taken from the conventional Skyrme Hartree--Fock calculations~\citep{Reinhard:1986qq,Sugahara:1993wz}.
The $R_{\rm ch}$ is defined as
\begin{equation}
  R_{\rm ch}=\sqrt{R_{p}^{2}+(0.8783)^{2}},
\end{equation}
with $R_{p}$ being the point proton radius~\citep{Chen:2014sca}.

We see the linear correlation between $R_{\rm skin}^{48}$ and $R_{\rm skin}^{208}$ in the left panel of Fig.~\ref{fig:NST}.
In general, the larger $R_{\rm skin}^{48}$ and $R_{\rm skin}^{208}$ are obtained by the models with the larger $L$ (see Table~\ref{tab:matter}).
To explain the results from RCNP, $L$ should be small such as the OMEG3, BigApple, FSUGarnet, and IU-FSU.
In contrast, the DINO family is located far from the points calculated by the other RMF models.
As explained in~\citet{Reed:2023cap}, the DINO family expresses the large $K_{\rm sym}$ by means of the huge $\delta$--$N$ and $\rho$--$N$ couplings.
Although it is difficult to support the PREX-2 and CREX results simultaneously, only the DINOc successfully aligns with both data sets.
\RevB{We note that the $\delta$-$N$ coupling and $\sigma$-$\delta$ mixing affect the charge radii of finite nuclei and hence $R_{\rm skin}$ while they  have less influence on the binding energy because we focus on the finite, closed-shell nuclei, $^{16}$O, $^{40,48}$Ca, $^{68}$Ni, $^{90}$Zr, $^{100,116,132}$Sn, and $^{208}$Pb, in the present study~\citep{Liliani:2023cfv}.}

To clarify the effect of $\delta$ meson on the characteristics of finite nuclei, we describe the correlation between $R_{\rm skin}^{48}$ and $R_{\rm skin}^{208}$ for the FSUGarnet, TAMUC-FSUa, and FSUGold2 series in the right panel of Fig.~\ref{fig:NST}.
We also display the calculations based on the other RMF models including the $\delta$ meson as well as the $\sigma$, $\omega$, and $\rho$ mesons.
As shown in Table~\ref{tab:CCs2}, $K_{\rm sym}$ becomes large as $g_{\delta}^{2}$ increases.
Consequently, the TAMUC-FSUa, and FSUGold2 series draw the lines from the upper right to the bottom left.
In particular, the FSUGold2 with the large $\delta$--$N$ coupling ($g_{\delta}^{2}\ge250$) supports both experimental data from the parity-violating electron scattering.
On the other hand, the FSUGarnet series moves away from the PREX-2 and CREX results when the large $g_{\delta}^{2}$ is introduced.

\begin{figure}[t!]
  \begin{center}
    \includegraphics[width=8.8cm]{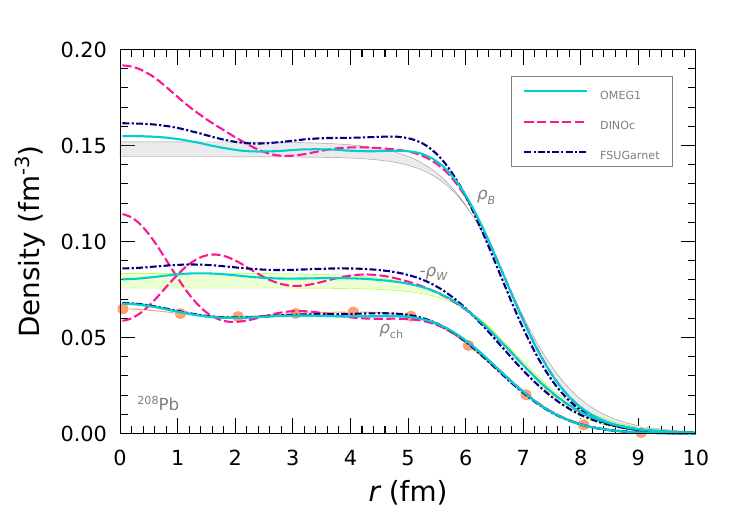}
    \includegraphics[width=8.8cm]{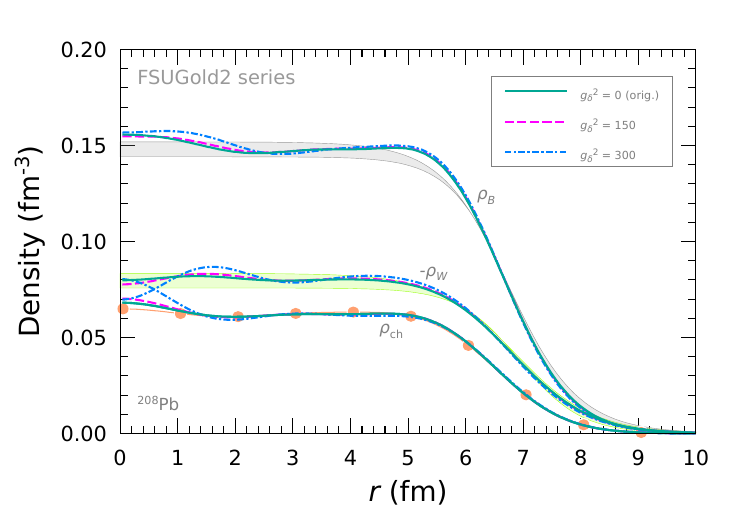}
  \end{center}
  \caption{Baryon, charge, and weak change densities, $\rho_{B}$, $\rho_{\rm ch}$, and $\rho_{W}$, for $^{208}$Pb.
    The density profiles for the OMEG1, DINOc, and FSUGarnet are given in the left panel.
    The right panel is for the FSUGold2 series.}
  \label{fig:density-profile}
\end{figure}
The density profiles in $^{208}$Pb are displayed in Fig.~\ref{fig:density-profile}.
We here present the baryon, charge, and weak charge densities, $\rho_{B}$~$(=\rho_{p}+\rho_{n})$, $\rho_{\rm ch}$, and $\rho_{W}$, with the experimental results~\citep{PREX:2021umo,DeVries:1987atn}.
The $\rho_{W}$ is approximately expressed as
\begin{equation}
  \rho_{W}\left(\bm{r}\right)
  \simeq Q_{p}\rho_{\rm ch}\left(\bm{r}\right)
  +      Q_{n}\int d\bm{r}^{\prime} \, G_{p}^{E}\left(\left|\bm{r}-\bm{r}^{\prime}\right|\right)\rho_{n}\left(\bm{r}\right),
\end{equation}
with $Q_{p(n)}$ being the proton (neutron) weak charge and $G_{p}^{E}$ being the proton electric form factor~\citep{Horowitz:1999fk,Horowitz:2012tj,Niksic:2002yp}.
The OMEG family is calibrated so as to reproduce $-\rho_{W}$ and $\rho_{B}$ in $^{208}$Pb by the PREX-2 experiment.

In the left panel of Fig.~\ref{fig:density-profile}, we present the density profiles for the OMEG1, DINOc, FSUGarnet.
The OMEG1 and FSUGarnet adequately satisfy the density distributions of $\rho_{\rm ch}$ from the elastic electron scattering~\citep{DeVries:1987atn}.
On the other hand, the DINOc possesses the instability around the core of nuclei because of the strong $\delta$--$N$ coupling constant~\citep{Reed:2023cap}.
As a result, the density profiles, $\rho_{B}$, $\rho_{\rm ch}$, and $\rho_{W}$, show the large density fluctuations around the core.

The effect of $\delta$--$N$ coupling on the density profiles for the FSUGold2 series is illustrated in the right panel of Fig.~\ref{fig:density-profile}.
There is almost no difference up to $g_{\delta}^{2}=150$.
In the case of $g_{\delta}^{2}=300$, $\rho_{\rm ch}$ and $\rho_{W}$ begin to show the instability around the core, but $\rho_{B}$ still matches the experimental data from PREX-2~\citep{PREX:2021umo}.
When the larger value, $g_{\delta}^{2}>300$, is taken, the unexpectedly large fluctuations of $\rho_{\rm ch}$ and $\rho_{W}$ emerge around the core, and the wave functions do not converge numerically.
In the present study, we thus impose the limit on the $\delta$--$N$ coupling as $g_{\delta}^{2}\le300$ for the FSUGold2 series.
We here comment that this defect can not be solved even if one considers the quartic self-interactions of $\delta$ and/or $\rho$ mesons in Eq.~\eqref{eq:NL-pot}, which less affect $R_{\rm skin}^{48}$ and $R_{\rm skin}^{208}$.

% \clearpage

\subsection{Infinite nuclear matter}

\begin{figure}[t!]
  \begin{center}
    \includegraphics[width=8.8cm]{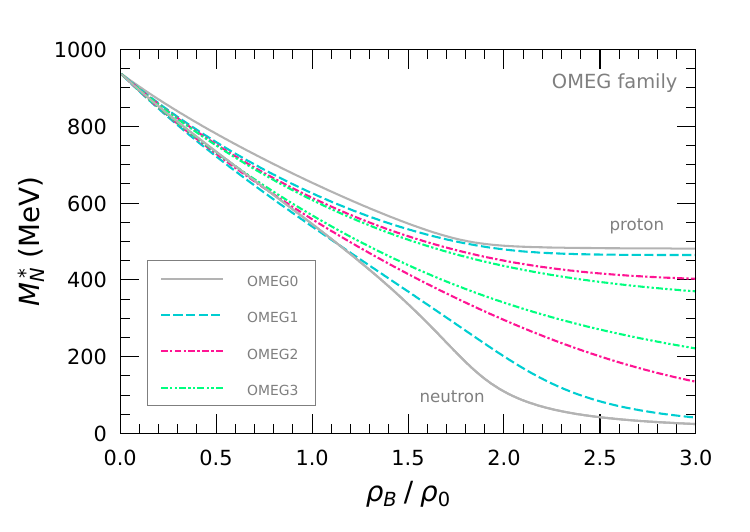}
    \includegraphics[width=8.8cm]{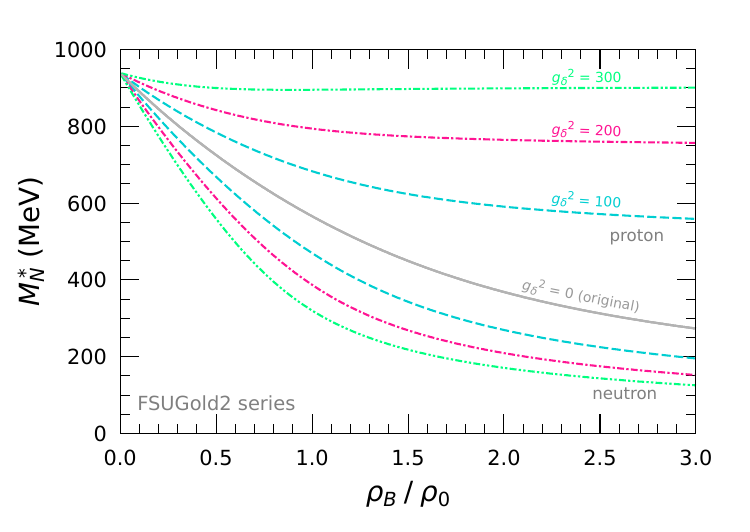}
  \end{center}
  \caption{Effective nucleon mass, $M_{N}^{\ast}$, as a function of $\rho_{B}/\rho_{0}$ for the OMEG family (left panel) and the FSUGold2 series (right panel).}
  \label{fig:Emass}
\end{figure}
The $\delta$-meson effect can be clearly seen in the effective nucleon mass, $M_{N}^{\ast}$, in Eq.~\eqref{eq:emass}.
Displayed in Fig.~\ref{fig:Emass} is the density dependence of $M_{N}^{\ast}$ in pure neutron matter for the OMEG family and the FSUGold2 series.
When the $\rho$ meson only is included, the RMF model gives the equal effective mass of proton and neutron.
However, the iso-scalar $\delta$ meson is responsible for the mass splitting between protons and neutrons, where $M_{p}^{\ast}$ is much heavier than $M_{n}^{\ast}$ at high densities.
Compared with the OMEG family, the FSUGold2 series shows the strong mass splitting, as $g_{\delta}^{2}$ increases, even at low densities.
It is implied that the neutron distribution is more spread out than the proton one, because $M_{n}^{\ast}$ is lighter, and then, the large fluctuations of $\rho_{\rm ch}$ and $\rho_{W}$ appear around the core of $^{208}$Pb as shown in Fig~\ref{fig:density-profile}.
% In contrast, the $\sigma$--$\delta$ mixing affects the mass splitting at high densities in the OMEG family.
% Especially, $M_{n}^{\ast}$ for the OMEG0 suddenly decreases above $1.5\rho_{0}$.
Due to the $\delta$--$N$ coupling and the $\sigma$--$\delta$ mixing, $M_{p}^{\ast}$ and $M_{n}^{\ast}$ respectively reach the almost constant values at high densities in all the cases.

\begin{figure}[t!]
  \begin{center}
    \includegraphics[width=12.0cm]{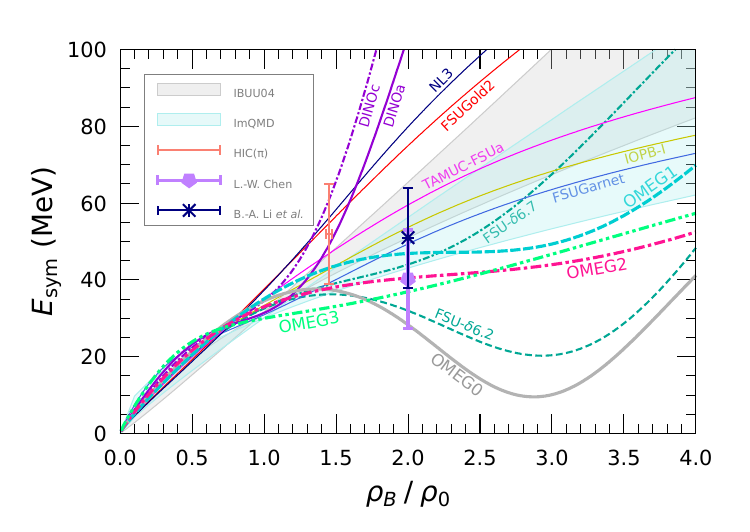}
  \end{center}
  \caption{Density dependence of nuclear symmetry energy, $E_{\rm sym}$.
    The shaded regions are the results from analyses of HIC data using the isospin-dependent Boltzmann-Uehling-Uhlenbec (IBUU04) and improved quantum molecular dynamics (ImQMD) transport models~\citep{Chen:2004si,Li:2005jy,Tsang:2008fd}.
    The recent experimental constraint from the pion emission in heavy-ion reactions is expressed as HIC($\pi$) with $E_{\rm sym}(\rho_{B})=52\pm13$~MeV at $\rho_{B}/\rho_{0}=1.45\pm0.2$~\citep{SpiRIT:2021gtq,Lynch:2021xkq,Tsang:2023vhh}.
    We also present two theoretical constraints on the magnitude of $E_{\rm sym}$ at $2\rho_{0}$ with $E_{\rm sym}(2\rho_{0})\simeq40.2\pm12.8$~MeV by \citet{Chen:2015gba} and $E_{\rm sym}(2\rho_{0})\simeq51\pm13$~MeV by \citet{Li:2021thg}.}
  \label{fig:Esym}
\end{figure}
The density dependence of nuclear symmetry energy, $E_{\rm sym}$, in Eq.~\eqref{eq:Esym-RMF} is depicted in Fig.~\ref{fig:Esym}.
We here present the calculations using the OMEG, FSU-$\delta$, and DINO families.
Furthermore, we use the conventional ones (the NL3, FSUGold2, TAMUC-FSUa, IOPB-I, and FSUGarnet models).
In addition, several experimental or theoretical constraints are presented.
Figure~\ref{fig:Esym} highlights significant differences in $E_{\rm sym}$ at high densities, that is, whereas the conventional calculations show a monotonic increase in $E_{\rm sym}$, the models with the $\delta$ meson exhibit more complex behavior.
In particular, the DINO family predicts a large $E_{\rm sym}$ above $1.5\rho_{0}$ as the $\delta$ meson amplifies $E_{\rm sym}$ in dense nuclear matter~\citep{Miyatsu:2022wuy}.
The OMEG and FSU-$\delta$ families, on the other hand, display unusual $E_{\rm sym}$ trends depending on the strength of $\delta$--$N$ coupling and $\sigma$--$\delta$ mixing.
The $\sigma$--$\delta$ mixing has a weak influence on $E_{\rm sym}$ below $\rho_{0}$, but, as discussed by \citet{Zabari:2018tjk}, it becomes substantial above $\rho_{0}$.
Specifically, the $\sigma$--$\delta$ mixing reduces $E_{\rm sym}$ at high densities, partially offsetting the increase from the $\delta$--$N$ interaction.
Furthermore, in the OMEG0 and FSU-$\delta6.2$, the inflection points appear above $\rho_{0}$ and the dip emerges around $2.5\rho_{0}$--$3.5\rho_{0}$.
This behavior is similar to the cusp in $E_{\rm sym}$ in the skyrmion crystal approach~\citep{Ma:2021nuf,Lee:2021hrw} and to the results from the Skyrme Hartree-Fock calculations~\citep{Chen:2005ti}.
We note that, as explained in Sec.~\ref{sec:stability}, the thermodynamic constraint on chemical potential in isospin-asymmetric nuclear matter, $V_{\mu}(\rho_{B},\alpha)>0$, is satisfied over all densities, namely $E_{\rm sym}(\rho_{B},\alpha)>E_{\rm sym}(\rho_{B})>0$.

\begin{figure}[t!]
  \begin{center}
    \includegraphics[width=8.8cm]{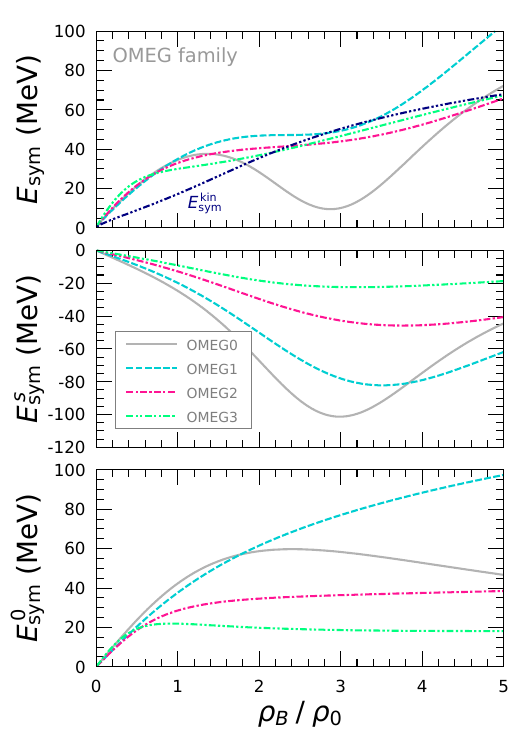}
    \includegraphics[width=8.8cm]{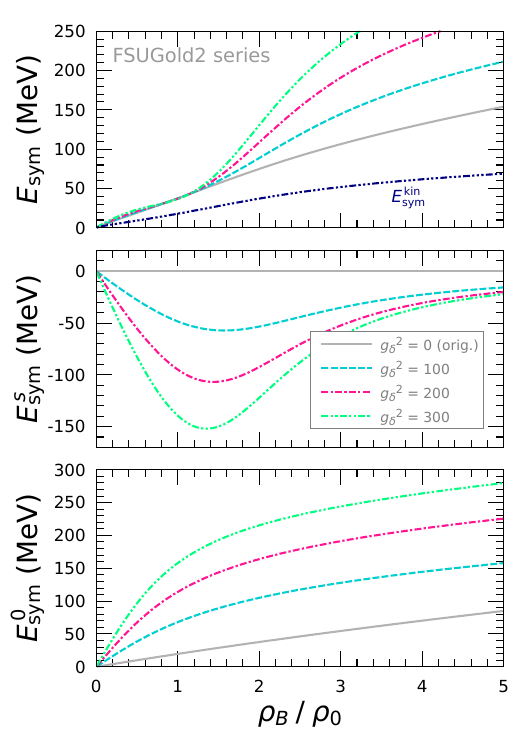}
  \end{center}
  \caption{Lorentz decomposition of nuclear symmetry energy, $E_{\rm sym}$, for the OMEG family (left panels) and the FSUGold2 series (right panels).
    The total $E_{\rm sym}$ and the kinetic term, $E_{\rm sym}^{\rm kin}$, are presented in the top panels.
    The scalar (time) component of potential term, $E_{\rm sym}^{s}$ $(E_{\rm sym}^{0})$, is given in the middle (bottom) panels.}
  \label{fig:Esym2}
\end{figure}
Based on the Lorentz decomposition of nucleon self-energy in Sec.\ref{sec:bulk}, $E_{\rm sym}$ is generally divided into the kinetic and potential terms, $E_{\rm sym}^{\rm kin}$ and $E_{\rm sym}^{\rm pot}$, as $E_{\rm sym}=E_{\rm sym}^{\rm kin}+E_{\rm sym}^{\rm pot}$.
In RMF approximation, only the isovector mesons contribute to $E_{\rm sym}^{\rm pot}$ as $E_{\rm sym}^{\rm pot}=E_{\rm sym}^{s}+E_{\rm sym}^{0}$, where the scalar ($s$) and time ($0$) components, $E_{\rm sym}^{s}$ and $E_{\rm sym}^{0}$, are respectively given by the $\delta$ and $\rho$ mesons.
We show the Lorentz decomposition of $E_{\rm sym}$ for the OMEG family and the FSUGold2 series as a function of $\rho_{B}/\rho_{0}$ in Fig.~\ref{fig:Esym2}.
The top panels are the density dependence of $E_{\rm sym}$ and $E_{\rm sym}^{\rm kin}$.
We see that the unique behavior of $E_{\rm sym}$ in the OMEG family is caused by $E_{\rm sym}^{\rm pot}$ because $E_{\rm sym}^{\rm kin}$ is almost the same as in both cases.
The contents of $E_{\rm sym}^{\rm pot}$ are given in the middle and bottom panels of Fig.~\ref{fig:Esym2}.
It is found that $E_{\rm sym}^{s}$ is negative while $E_{\rm sym}^{0}$ is positive, which is similar to the general understanding of $N$--$N$ interaction described by the nuclear attractive and repulsive forces.
\RevA{Note that a similar description of $E_{\rm sym}^{\rm pot}$ has been reported using the RMF model with a contact interaction of isovector mesons, where the scalar contribution, $(\bar{\psi}_{N}\bm{\tau}_{N}\psi_{N})^{2}$, is positive while the vector one, $(\bar{\psi}_{N}\gamma_{\mu}\bm{\tau}_{N}\psi_{N})^{2}$, is negative~\citep{Maruyama:1999td,Maruyama:2021ghf}.}
% There seems to be a large differences, namely how the $\delta$ meson affects $E_{\rm sym}$, between the OMEG families and the FSUGold2 series.
It is noticeable that, for the FSUGold2 series, $E_{\rm sym}^{s}$ is strongly influenced by the $\delta$--$N$ coupling above $\rho_{0}$, and the contribution of $E_{\rm sym}^{s}$ is small at high densities.
Conversely, for the OMEG family, the $\sigma$--$\delta$ mixing shows less impact on $E_{\rm sym}^{s}$ below $\rho_{0}$, but it strongly affects $E_{\rm sym}^{s}$ at high densities.
When the absolute value of $E_{\rm sym}^{s}$ is larger than that of $E_{\rm sym}^{0}$, $E_{\rm sym}^{\rm pot}$ has the rapid reduction, and then $E_{\rm sym}$ shows a dip around $3\rho_{0}$ as in the cases for the OMEG0 and FSU-$\delta6.2$ in Fig.~\ref{fig:Esym}.

\begin{figure}[t!]
  \begin{center}
    \includegraphics[width=11.0cm]{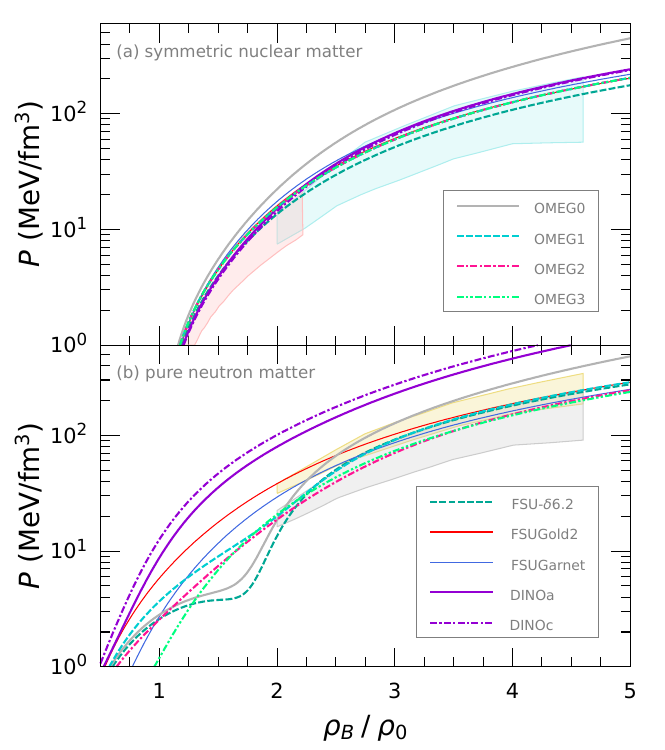}
  \end{center}
  \caption{EoS---pressure, $P$, as a function of $\rho_{B}/\rho_{0}$---for (a) symmetric nuclear matter and for (b) pure neutron matter.
    The shaded areas represent the constraints from elliptical flow data~\citep{Danielewicz:2002pu} and kaon production data~\citep{Fuchs:2005zg,Lynch:2009vc}.}
  \label{fig:Press}
\end{figure}
The EoSs for symmetric nuclear matter and pure neutron matter are displayed in Fig.~\ref{fig:Press} with the constraints on the nuclear EoS extracted from the analyses of particle flow data in HICs~\citep{Danielewicz:2002pu,Fuchs:2005zg,Lynch:2009vc}.
In both panels, we show the various EoSs calculated by the OMEG, DINO, and FSU-$\delta$ families, and the FSUGarnet and FSUGold2 models.
The $\delta$ meson does not affect $P$ in symmetric nuclear matter.
All the cases except for the OMEG0 are well constructed to match the HIC data in symmetric nuclear matter because of the small $K_{0}$.
However, the stiffer EoS with $K_{0}\simeq285$~MeV is still acceptable, taking into account the recent simulation of Au$+$Au collisions~\citep{Oliinychenko:2022uvy}.
In contrast, the $\delta$ meson has a large impact on $P$ in pure neutron matter.
The DINOa and DINOc show the hard EoSs, which are far from the constraints from HICs, due to the large $\delta$--$N$ coupling.
Meanwhile, the strong $\sigma$--$\delta$ mixing softens the EoSs extremely for the OMEG and FSU-$\delta$ families in the density region from $\rho_{0}$ to $2\rho_{0}$, around which the characteristics of a canonical $1.4~M_{\odot}$ neutron star are generally determined.

\begin{figure}[t!]
  \begin{center}
    \includegraphics[width=8.8cm]{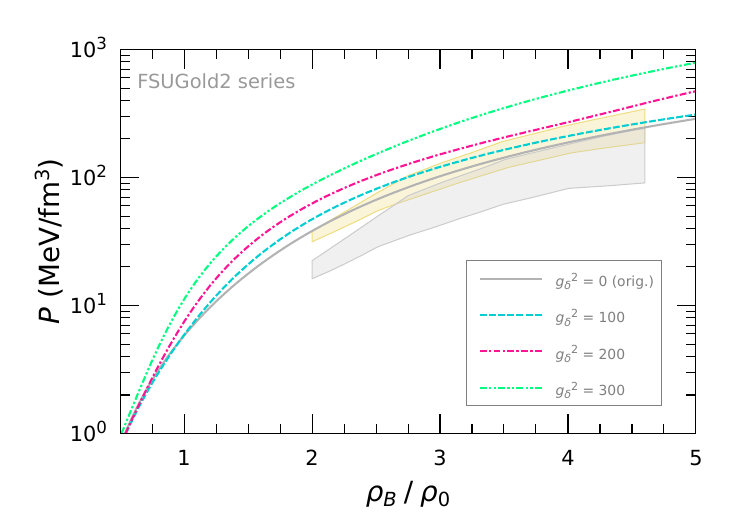}
    \includegraphics[width=8.8cm]{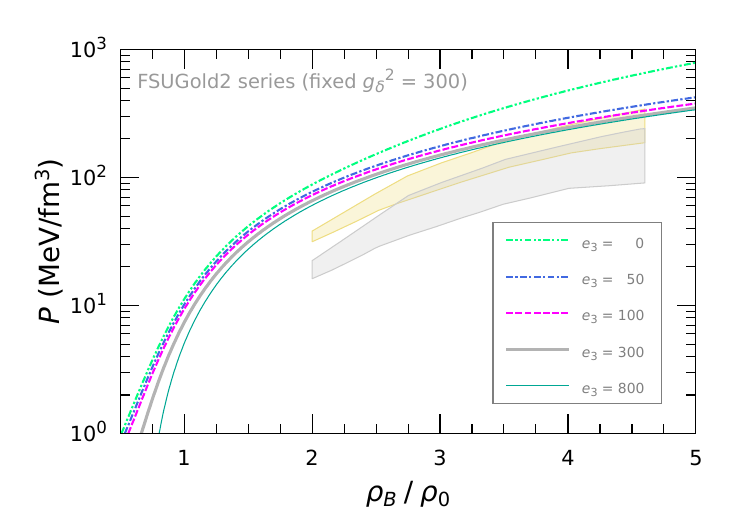}
  \end{center}
  \caption{EoS for pure neutron matter for the FSUGold2 series.
    The left panel shows the dependence of $\delta$--$N$ coupling square, $g_{\delta}^{2}$.
    In the right panel, the influence of quartic $\rho$-meson self-interaction, $e_{3}$, is presented with the fixed parameter of $g_{\delta}^{2}=300$ (see Table~\ref{tab:CCs3}).}
  \label{fig:Press-FSU2}
\end{figure}
We present the EoS for pure neutron matter for the FSUGold2 series in Fig.~\ref{fig:Press-FSU2}.
In the left panel, the EoS becomes hard with increasing the $\delta$--$N$ coupling, and the EoS with $g_{\delta}^{2}=300$ exceeds the HIC results as in the cases for the DINO family in Fig.~\ref{fig:Press}.
Hence, we find that, even if the large $\delta$--$N$ coupling is introduced simply, it is not easy to explain simultaneously both properties of dense nuclear matter and characteristics of finite nuclei for $R_{\rm skin}^{48}$ and $R_{\rm skin}^{208}$ in Fig.~\ref{fig:NST}.
In order to suppress such excessive stiffness of EoSs for pure neutron matter due to the $\delta$--$N$ coupling, we additionally include the quartic self-interaction of $\rho$ meson in the FSUGold2 model with the upper limit of $g_{\delta}^{2}=300$ (see Table~\ref{tab:CCs3}), given in the right panel of Fig.~\ref{fig:Press-FSU2}.
The EoS is soft and again reaches the upper edge of the constraint from HICs with increasing the quartic coupling, $e_{3}$, whose effect is almost imperceptible below $\rho_{0}$.

% \clearpage

\subsection{Neutron star physics}

\input{./Tables/Table-NS-all.tex}
In studying neutron star physics, the EoS for nonuniform matter is additionally required as well as that for uniform nuclear matter since the radius of a neutron star is remarkably sensitive to the nuclear EoS at very low densities~\citep{Glendenning:1997wn}.
In the present study, to cover the crust region, we adopt the MYN13 EoS, in which nuclei are taken into consideration using the Thomas-Fermi calculation in nonuniform matter and the EoS for infinite nuclear matter is constructed with the relativistic Hartree-Fock calculation~\citep{Miyatsu:2011bc,Katayama:2012ge,Miyatsu:2013hea,Miyatsu:2015kwa}.
We list in Table~\ref{tab:NS} the predicted stellar properties, which are calculated by solving the Tolman--Oppenheimer--Volkoff (TOV) equation~\citep{Tolman:1939jz,Oppenheimer:1939ne}.

There are three methods used widely to determine the crust-core transition density, $\rho_{t}$~\citep{Sulaksono:2014bja}: the thermo-dynamical method, the dynamical method, and the random-phase-approximation method.
We employ the first method in the present study.
As explained in Sec.~\ref{sec:stability}, the stability of nuclear and neutron star matter is determined by the constraints on chemical potential and pressure, $V_{\mu}(\rho_{B},\alpha)>0$ and $V_{P}(\rho_{B},\alpha)>0$, in the first law of thermodynamics.
Since the proton fraction, $Y_{p}$, is supposed to be small in the crust region, the second-order Taylor series approximation of the nuclear EoS is generally adopted in the density derivative of baryon pressure, $\left(\partial P_{B}/\partial\rho_{B}\right)_{\mu}$, in Eq.~\eqref{eq:chemical-potential-condition-baryon}~\citep{Kubis:2006kb}.
However, it has been reported that the parabolic approximation of isospin-asymmetric nuclear EoS may be misleading as regards the predictions for $\rho_{t}$~\citep{Routray:2016yvp}.
We thus employ the exact nuclear EoS to calculate $V_{P}$ defined in Eq.~\eqref{eq:thermodynamic-condition}.

\begin{figure}[t!]
  \begin{center}
    \includegraphics[width=12.0cm]{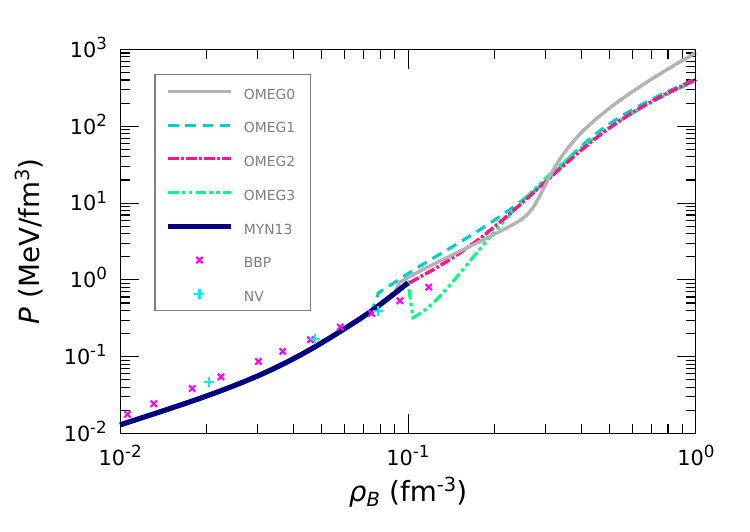}
  \end{center}
  \caption{EoS for neutron star matter for the OMEG family.
    The inner-crust region is described by the EoSs of MYN13~\citep{Miyatsu:2013hea}, BBP~\citep{Baym:1971ax}, and NV~\citep{Negele:1971vb}.}
  \label{fig:Phase}
\end{figure}
\begin{figure}[t!]
  \begin{center}
    \includegraphics[width=8.8cm]{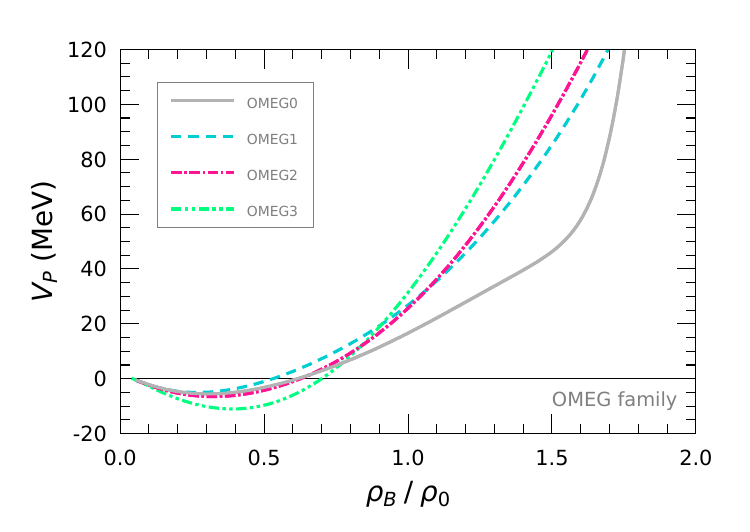}
    \includegraphics[width=8.8cm]{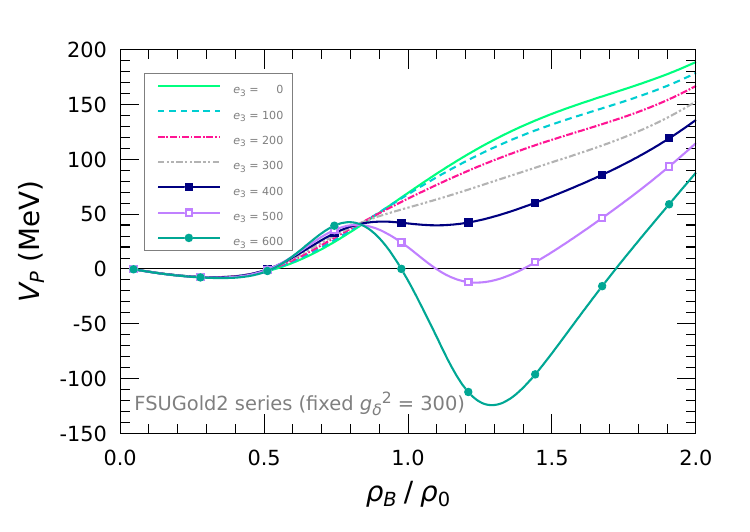}
  \end{center}
  \caption{Thermodynamic stability of pressure, $V_{p}$, in neutron star matter for the OMEG family (left panel) and for the FSUGold2 series with the fixed parameter of $g_{\delta}^{2}=300$ (right panel).}
  \label{fig:Stability}
\end{figure}
We summarize the results of $\rho_{t}$ in the second column of Table~\ref{tab:NS}.
Compared with the results from the Taylor series expansion, the our results settle between the second-order and fourth-order calculations.
For example, for the FSUGold, the exact value is $\rho_{t}=0.079$ fm$^{-3}$ while the second-order (fourth-order) result is $\rho_{t}^{2\rm nd}=0.089$ ($\rho_{t}^{4\rm th}=0.051$) fm$^{-3}$ (see Table~2 in~\citet{Routray:2016yvp}).
In addition, the current results are almost the same as the transition density from the pasta phase to the homogeneous nuclear matter in the model calculation with Thomas-Fermi approximation~\citep{Providencia:2013dsa}.
The EoS for neutron star matter in the OMEG family is presented in Fig.~\ref{fig:Phase}.
The crust-core phase transition occurs at $V_{P}=0$, which is also described in the left panel of Fig.~\ref{fig:Stability}.
As it is well known, the EoS with larger $L$ gives the smaller $\rho_{t}$~\citep{Li:2019xxz}.

Since the large $\sigma$--$\delta$ mixing enhances the rapid reduction of $E_{\rm sym}$ around $3\rho_{0}$ as shown in Fig.~\ref{fig:Esym}, we have to investigate the stability of neutron star matter.
Similar to the crust-core phase transition, we adopt the thermo-dynamical method.
It is especially important to apply the exact nuclear EoS to $V_{P}$ because $Y_{p}$ is by no means small and the Taylor series expansion is prohibited at high densities.
It is found that the constraint on chemical potential, $V_{\mu}>0$, is always satisfied as $E_{\rm sym}$ is positive at any densities.
Hence, all we have to do is check the thermodynamic stability of pressure, $V_{P}$.
In Fig.~\ref{fig:Stability}, we show $V_{P}$ in neutron star matter.
In general, $V_{P}$ changes from negative to positive at $\rho_{t}$, and the stable EoS possesses $V_{P}>0$ even at high densities.
Despite the OMEG0 give a strong concavity in $E_{\rm sym}$ by the $\sigma$--$\delta$ mixing, it satisfies the thermodynamic stability.
In the right panel of Fig.~\ref{fig:Stability}, we show $V_{P}$ for the FSUGold2 series.
The neutron star matter keeps $V_{P}>0$ when the $\delta$--$N$ coupling only is included, whereas the large quartic self-interaction of $\rho$ meson, $e_{3}$, makes the matter unstable.
Though the quartic $\rho$-meson self-interaction is useful to figure out the HIC data as mentioned in the right panel of Fig.~\ref{fig:Press-FSU2}, the large value of $e_{3}$ is unfavorable to the neutron star physics.

\begin{figure}[t!]
  \begin{center}
    \includegraphics[width=12.0cm]{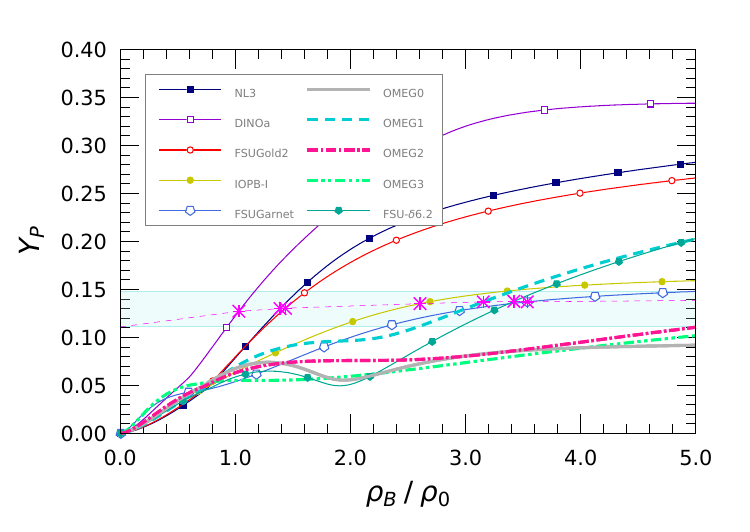}
  \end{center}
  \caption{Proton fraction, $Y_{p}$ $(=\rho_{p}/\rho_{B})$, in neutron star matter.
    The shaded band is the threshold for the direct URCA process~\citep{Maruyama:1999td,Horowitz:2002mb}.
    \RevA{The asterisks indicate the densities at which the direct URCA process actually begins.}}
  \label{fig:Yp}
\end{figure}
We illustrate in Fig.~\ref{fig:Yp} the proton fraction, $Y_{p}$, in neutron star matter with the threshold for the direct URCA process.
The direct URCA process is visible only when $Y_{p}$ is large enough to conserve momentum in $\beta$-equilibrated matter, in which the Fermi momenta of neutrons, protons, and electron must satisfy the relation: $k_{F_{n}}\le k_{F_{p}}+k_{F_{e}}$.
Hence, $Y_{p}$ can be estimated as $0.111\le Y_{p}\le0.148$, above which the direct URCA cooling occurs~\citep{Maruyama:1999td,Horowitz:2002mb,Page:2005fq}.
\RevA{We find that as $\rho_{B}$ increases, the threshold of $Y_{p}$ for the direct URCA process shifts toward the upper boundary where muons are present.}
The $Y_{p}$ for the DINOa grows quickly with increasing $\rho_{B}$ due to the large $\delta$--$N$ coupling, and then the direct URCA process is allowed sufficiently at $2\rho_{0}$, which corresponds to the core density of a canonical $1.4~M_{\odot}$ neutron star.
Conversely, in the OMEG and FSU-$\delta$ families, the $\sigma$--$\delta$ mixing suppresses $Y_{p}$, and then delays the direct URCA process.
Particularly, the direct URCA process never occurs for the OMEG0, OMEG2, and OMEG3 in the current density region, and thus the so-called modified URCA process, which is the standard model of neutron-star coolings, mainly takes place for the neutrino emission \citep{Lattimer:1991ib}.
Alternatively, the possibility of exotic degrees of freedom in the core of a neutron star, such as hyperons, quarks, gluons and/or some unusual condensations of boson-like matter, should be taken into account to understand the rapid neutron star cooling.

\begin{figure}[t!]
  \begin{center}
    \includegraphics[width=12.0cm]{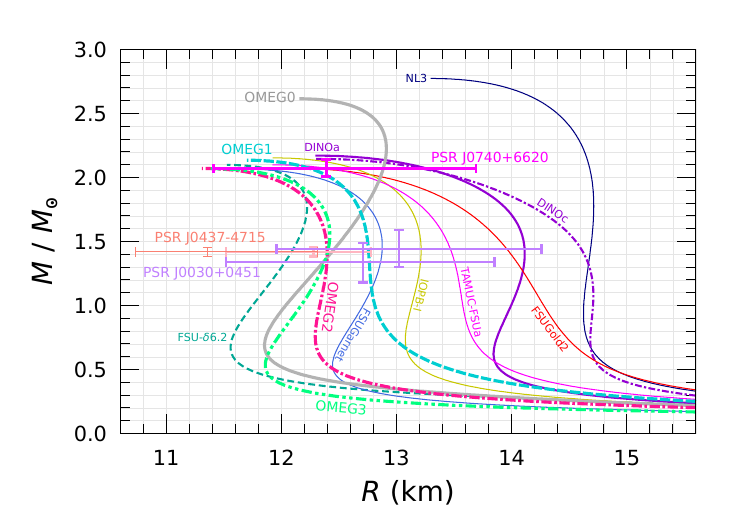}
  \end{center}
  \caption{Mass--radius relations of neutron stars.
    The NICER observation data are supplemented by the constraints from PSR J0030$+$0451 ($1.44_{-0.14}^{+0.15}$~$M_{\odot}$ and $13.02_{-1.06}^{+1.24}$~km, and $1.34_{-0.16}^{+0.15}$~$M_{\odot}$ and $12.71_{-1.19}^{+1.14}$~km)~\citep{Miller:2019cac,Riley:2019yda}, PSR J0740$+$6620 ($2.072^{+0.067}_{-0.066}$~$M_{\odot}$ and $12.39^{+1.30}_{-0.98}$~km)~\citep{NANOGrav:2019jur,Fonseca:2021wxt,Riley:2021pdl}, and PSR J0437$–$4715 ($12.28^{+0.50}_{-0.76}$~km at $1.4$~$M_{\odot}$, and $1.418\pm0.037$~$M_{\odot}$ and $11.36^{+0.95}_{-0.63}$~km)~\citep{Rutherford:2024srk,Choudhury:2024xbk}.
    % We also give the original information on a central compact object (CCO) within the supernova remnant (SNR), HESS J1731$-$347 ($0.83_{-0.13}^{+0.17}$~$M_{\odot}$ and $11.25_{-0.37}^{+0.53}$~km)~\citep{2022NatAs...6.1444D}.
  }
  \label{fig:MR}
\end{figure}
The mass($M$)--radius($R$) relations of neutron stars are displayed in Fig.~\ref{fig:MR}.
We here show the astrophysical constraints from the NICER observations: PSR J0030$+$0451~\citep{Miller:2019cac,Riley:2019yda}, PSR J0740$+$6620~\citep{NANOGrav:2019jur,Fonseca:2021wxt,Riley:2021pdl}, and PSR J0437$–$4715~\citep{Rutherford:2024srk,Choudhury:2024xbk}.
According to the observation from PSR J0740$+$6620, the maximum mass of a neutron star, $M_{\rm max}$, should be larger than $2M_{\odot}$.
Thus the EoS involving the large $R$, such as the NL3, is ruled out.
It is found that the large $\delta$--$N$ coupling affects the large $R$ in the DINO family, whereas the $\sigma$--$\delta$ mixing makes $R$ small in the OMEG family.
In particular, though the DINOa and OMEG0 have the same $L$ as $L=50$~MeV, their $M$--$R$ relations are completely different and the difference of $R$ at canonical-mass point reads approximately $1.7$~km (see also Table~\ref{tab:NS}).
The OMEG family can support not only the NICER constraint on $R_{1.4}$ from PSR J0030$+$0451 but also that from PSR J0437$–$4715, which is the latest result based on new chiral effective field theory inputs~\citep{Rutherford:2024srk}.

\begin{figure}[t!]
  \begin{center}
    \includegraphics[width=12.0cm]{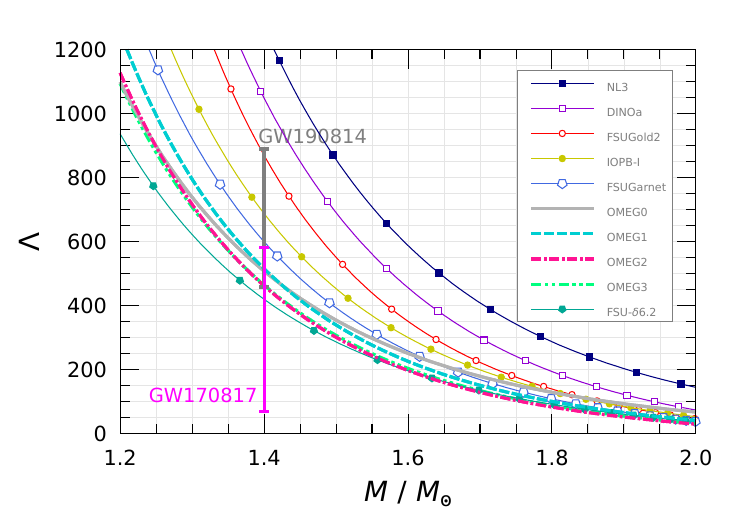}
  \end{center}
  \caption{Dimensionless tidal deformability, $\Lambda$, of neutron stars.
    We present the constraints on $\Lambda_{1.4}$ from the binary merger events, GW170817 ($\Lambda_{1.4}=190^{+390}_{-120}$)~\citep{LIGOScientific:2018cki} and GW190814 ($\Lambda_{1.4}=616^{+273}_{-158}$)~\citep{LIGOScientific:2020zkf}.}
  \label{fig:Tidal}
\end{figure}
The dimensionless tidal deformability, $\Lambda$, of neutron stars is displayed in Fig.~\ref{fig:Tidal} as a function of $M/M_{\odot}$.
The $\Lambda$ is defined as $\Lambda = \frac{2}{3}k_{2}\left(R/M\right)^{5}$ with $k_{2}$ being the second Love number~\citep{Hinderer:2007mb,Hinderer:2009ca}.
The astrophysical constraints on $\Lambda$ at the canonical-mass point, $\Lambda_{1.4}$, from the binary merger events detected by the Advanced LIGO and Advanced Virgo observatories are also presented as follows: $\Lambda_{1.4}=190^{+390}_{-120}$ for GW170817~\citep{LIGOScientific:2018cki} and $\Lambda_{1.4}=616^{+273}_{-158}$ for GW190814~\citep{LIGOScientific:2020zkf}.
As explained in \citet{Miyatsu:2022wuy}, the $\delta$--$N$ coupling enlarges $\Lambda$ for the DINOa, and then $\Lambda_{1.4}$ lies far from the constraints on $\Lambda_{1.4}$ from GW190814.
On the other hand, the $\sigma$--$\delta$ mixing has a promising effect on $\Lambda$, and thus the OMEG family sufficiently matches the severe constraints from both GW170817 and GW190814.

% \clearpage

\section{Summary and conclusion} \label{sec:summary}

We have developed a new family of nuclear EoSs, referred to as the OMEG family, using the RMF model with nonlinear couplings between the isoscalar and isovector mesons.
In addition to the $\sigma$, $\omega$, and $\rho$ mesons, we have also included the $\delta$ meson to examine the ground-state properties of finite, closed-shell nuclei as well as the characteristics of nuclear and neutron star matter.
Specifically, we have investigated the effects of $\delta$--$N$ coupling and $\sigma$--$\delta$ mixing on the EoS for both nuclear and neutron star matter.
The model parameters for the OMEG family have been calibrated so as to satisfy the constraints from the particle flow data in HICs~\citep{Danielewicz:2002pu,Fuchs:2005zg,Lynch:2009vc}, the observed neutron-star mass of PSR J0740$+$6620~\citep{NANOGrav:2019jur,Fonseca:2021wxt,Riley:2021pdl}, and the dimensionless tidal deformability, $\Lambda_{1.4}$, from the neutron star merger, GW170817~\citep{LIGOScientific:2018cki}, as well as the results from the PREX-2 and CREX experiments~\citep{PREX:2021umo,CREX:2022kgg}.

It has been found that the $\delta$--$N$ coupling and the $\sigma$--$\delta$ mixing significantly influence the properties of isospin-asymmetric nuclear matter and finite nuclei, playing a crucial role in reconciling terrestrial experiments with astrophysical observations of neutron stars.
The strong $\delta$--$N$ coupling for the FSUGold2 series can simultaneously explain the large $R_{\rm skin}^{208}$ and the small $R_{\rm skin}^{48}$ measured by the PREX-2 and CREX experiments.
However, it seems difficult that the FSUGold2 series satisfy the combined constraints from the particle flow data in HICs and astrophysical observations, such as the EoS for pure neutron matter and the $\Lambda$ of neutron stars.
Even with the inclusion of quartic $\rho$-meson self-interaction in the FSUGold2 series, both experimental and observational results can not be understood, because the large $e_{3}$ destabilizes neutron star matter.
In contrast, the OMEG family can satisfy the recent measurement of $R_{1.4}=12.28^{+0.50}_{-0.76}$~km for PSR J0437$–$4715 from NICER~\citep{Rutherford:2024srk} and the stringent constraint on $\Lambda_{1.4}=190^{+390}_{-120}$ from GW170817~\citep{LIGOScientific:2018cki}.
This is attributed to the $\sigma$--$\delta$ mixing, which suppresses $E_{\rm sym}$ above $2\rho_{0}$, resulting in a softer nuclear EoS in the density region corresponding to the core density of the canonical neutron stars.

\RevB{In a future work, we plan to extend the present study to global calculations of finite nuclei properties covered the periodic table, aiming to achieve well-calibrated parameter sets for the RMF models.}
Finally, we comment that the further theoretical studies are necessary to reconcile the $R_{\rm skin}$ measured by proton (in)elastic scattering with that obtained from parity-violating electron scattering.
In particular, it is very significant to investigate the discrepancy between the PREX-2 data~\citep{PREX:2021umo} and the results from RCNP~\citep{Zenihiro:2010zz,Tamii:2011pv} and MAMI~\citep{Tarbert:2013jze}.
It is also essential to consider the effect of isospin symmetry breaking on asymmetric nuclear matter from the quark level~\citep{Guichon:1987jp,Saito:1994ki,Saito:1994tq,Saito:2005rv,Saito:2024qth,Nagai:2008ai}.

% \section*{Appendix} \label{sec:appendix}

% \section{Manuscript Formatting}

% \subsection{Heading Levels}

% %There are 5 heading levels

% \subsection{Level 2}
% \subsubsection{Level 3}
% \paragraph{Level 4}
% \subparagraph{Level 5}

% \subsection{Equations}
% Equations should be inserted in editable format from the equation editor.

% \begin{equation}
% \sum x+ y =Z\label{eq:01}
% \end{equation}

% \section{Additional Requirements}

% For additional requirements for specific article types and further information please refer to \href{http://www.frontiersin.org/about/AuthorGuidelines#AdditionalRequirements}{Author Guidelines}.

% \section*{Conflict of Interest Statement}
% %All financial, commercial or other relationships that might be perceived by the academic community as representing a potential conflict of interest must be disclosed. If no such relationship exists, authors will be asked to confirm the following statement: 

% The authors declare that the research was conducted in the absence of any commercial or financial relationships that could be construed as a potential conflict of interest.

% \section*{Author Contributions}

% The Author Contributions section is mandatory for all articles, including articles by sole authors. If an appropriate statement is not provided on submission, a standard one will be inserted during the production process. The Author Contributions statement must describe the contributions of individual authors referred to by their initials and, in doing so, all authors agree to be accountable for the content of the work. Please see  \href{https://www.frontiersin.org/about/policies-and-publication-ethics#AuthorshipAuthorResponsibilities}{here} for full authorship criteria.

\section*{Acknowledgments}
% This is a short text to acknowledge the contributions of specific colleagues, institutions, or agencies that aided the efforts of the authors.
T.M. would like to thank H.~Sagawa and G.~Col\`o for informative discussions of the neutron skin thickness of heavy nuclei.

\section*{Funding}
% Details of all funding sources should be provided, including grant numbers if applicable. Please ensure to add all necessary funding information, as after publication this is no longer possible.
This work was supported by the National Research Foundation of Korea (Grant Nos. RS-2023-00242196, NRF-2021R1A6A1A03043957, NRF-2020R1A2C3006177, and NRF-2018R1A5A1025563).

% \appendix

% \section*{Appendix A: Tables of model parameters and saturation properties} \label{sec:appendix-A}

\bibliographystyle{Frontiers-Harvard} %  Many Frontiers journals use the Harvard referencing system (Author-date), to find the style and resources for the journal you are submitting to: https://zendesk.frontiersin.org/hc/en-us/articles/360017860337-Frontiers-Reference-Styles-by-Journal. For Humanities and Social Sciences articles please include page numbers in the in-text citations
% \bibliographystyle{Frontiers-Vancouver} % Many Frontiers journals use the numbered referencing system, to find the style and resources for the journal you are submitting to: https://zendesk.frontiersin.org/hc/en-us/articles/360017860337-Frontiers-Reference-Styles-by-Journal
% \bibliography{test}

\bibliography{Nskin}

%%% Make sure to upload the bib file along with the tex file and PDF
%%% Please see the test.bib file for some examples of references

\end{document}

%% file: Tables/Table-CCs.tex
\begin{table}[b!]
  \centering
  \caption{\label{tab:CCs}
    Model parameters for various effective interactions.
    The nucleon and $\delta$-meson masses in free space are fixed at $M_{N}=939$ MeV and $m_{\delta}=980$ MeV.
    For the HPNL0 and HPNL5, we set $m_{\delta}=983$ MeV~\citep{PhysRevC.108.055802}.
    The $M_{N}$ is taken as $M_{N}=938$ MeV only for the TM1~\citep{Sugahara:1993wz}.}
  \resizebox{\textwidth}{!}{
  % \rotatebox{90}{
    \begin{tabular}{lrrrrrrrrrrrr}
      \toprule
      \               & $m_{\sigma}$ & $m_{\omega}$ & $m_{\rho}$ & $g_{\sigma}^{2}$ & $g_{\omega}^{2}$ & $g_{\delta}^{2}$ & $g_{\rho}^{2}$ & $g_{2}$     &   $g_{3}$ &   $c_{3}$ & $\Lambda_{\sigma\delta}$ & $\Lambda_{\omega\rho}$ \\
      Models          &        (MeV) &        (MeV) &      (MeV) &                \ &                 \ &               \ &              \ & (fm$^{-1}$) &         \ &      \  &                        \ &                        \ \\
      \midrule
      OMEG0           &    $496.500$ &    $782.660$ &  $775.260$ &         $89.384$ &        $142.847$ &         $37.699$ &       $51.744$ &     $9.976$ & $-21.469$ &        -- &                 $87.000$ &              $102.612$ \\
      OMEG1           &    $497.825$ &    $782.660$ &  $775.260$ &         $99.645$ &        $166.268$ &         $30.000$ &       $44.591$ &     $7.824$ &  $-1.115$ & $100.000$ &                 $95.000$ &               $75.677$ \\
      OMEG2           &    $497.820$ &    $782.660$ &  $775.260$ &         $99.641$ &        $166.269$ &         $20.000$ &       $44.364$ &     $7.823$ &  $-1.113$ & $100.000$ &                 $85.000$ &              $288.859$ \\
      OMEG3           &    $498.015$ &    $782.660$ &  $775.260$ &         $99.713$ &        $166.272$ &         $15.000$ &       $57.550$ &     $7.827$ &  $-1.105$ & $100.000$ &                 $70.000$ &              $909.825$ \\
      BigApple        &    $492.730$ &    $782.500$ &  $763.000$ &         $93.507$ &        $151.684$ &               -- &       $50.139$ &    $11.944$ & $-31.832$ &   $2.684$ &                       -- &             $1449.033$ \\
      DINOa           &    $490.050$ &    $782.500$ &  $763.000$ &         $93.942$ &        $154.443$ &        $278.788$ &      $201.454$ &    $11.501$ & $-20.139$ &  $59.626$ &                       -- &              $205.236$ \\
      DINOb           &    $485.795$ &    $782.500$ &  $763.000$ &         $91.032$ &        $150.806$ &        $313.178$ &      $219.270$ &    $11.651$ & $-20.561$ &  $56.870$ &                       -- &              $185.311$ \\
      DINOc           &    $484.162$ &    $782.500$ &  $763.000$ &         $90.648$ &        $151.032$ &        $335.813$ &      $230.653$ &    $11.642$ & $-20.517$ &  $57.039$ &                       -- &              $171.604$ \\
      FSU-$\delta$6.2 &    $491.500$ &    $782.500$ &  $763.000$ &        $104.332$ &        $180.507$ &         $38.440$ &       $48.867$ &     $8.090$ &   $5.881$ & $172.150$ &                $107.650$ &               $91.639$ \\
      FSU-$\delta$6.7 &    $491.500$ &    $782.500$ &  $763.000$ &        $104.332$ &        $180.507$ &         $44.890$ &       $52.843$ &     $8.090$ &   $5.881$ & $172.150$ &                 $90.157$ &              $102.004$ \\
      FSUGarnet       &    $496.939$ &    $782.500$ &  $763.000$ &        $110.350$ &        $187.693$ &               -- &       $47.966$ &     $9.565$ &  $-7.122$ & $137.981$ &                       -- &             $1555.729$ \\
      FSUGarnet+R \hspace*{-2.5cm}
                      &    $495.633$ &    $782.500$ &  $763.000$ &        $109.045$ &        $186.307$ &               -- &       $35.654$ &     $9.407$ &  $-6.452$ & $138.011$ &                       -- &             $1016.385$ \\
      FSUGold         &    $491.500$ &    $782.500$ &  $763.000$ &        $112.204$ &        $204.556$ &               -- &       $34.629$ &     $4.277$ &  $49.858$ & $418.394$ &                       -- &             $ 850.374$ \\
      FSUGold2        &    $497.479$ &    $782.500$ &  $763.000$ &        $108.070$ &        $183.733$ &               -- &       $20.145$ &     $8.546$ &  $-1.021$ & $144.122$ &                       -- &               $12.289$ \\
      FSUGold2+R \hspace*{-0.2cm}
                      &    $501.611$ &    $782.500$ &  $763.000$ &        $103.793$ &        $169.483$ &               -- &       $32.090$ &    $10.150$ & $-19.026$ &  $55.970$ &                       -- &              $687.024$ \\
      HPNL0           &    $501.825$ &    $782.500$ &  $770.000$ &        $109.108$ &        $180.637$ &          $1.443$ &       $26.967$ &     $9.393$ &  $-3.405$ & $134.110$ &                       -- &              $267.522$ \\
      HPNL5           &    $500.972$ &    $782.500$ &  $770.000$ &        $103.502$ &        $169.568$ &          $8.903$ &       $39.724$ &     $9.270$ &  $-2.866$ & $120.751$ &                 $46.077$ &              $511.487$ \\
      IOPB-I          &    $500.000$ &    $782.500$ &  $762.500$ &        $107.971$ &        $178.268$ &               -- &       $30.955$ &    $10.517$ & $-15.146$ &  $91.108$ &                       -- &              $528.819$ \\
      IU-FSU          &    $491.500$ &    $782.500$ &  $763.000$ &         $99.427$ &        $169.835$ &               -- &       $46.172$ &     $8.497$ &   $0.462$ & $144.219$ &                       -- &             $1442.856$ \\
      NL3             &    $508.194$ &    $782.501$ &  $763.000$ &        $104.408$ &        $165.562$ &               -- &       $19.883$ &    $10.454$ & $-28.955$ &        -- &                       -- &                     -- \\
      PD15            &    $480.025$ &    $780.000$ &  $763.000$ &         $97.535$ &        $173.792$ &         $22.782$ &       $60.551$ &     $7.662$ &   $6.551$ & $169.689$ &                       -- &              $627.406$ \\
      TAMUC-FSUa \hspace*{-3.5cm}
                      &    $502.200$ &    $782.500$ &  $763.000$ &        $106.504$ &        $176.178$ &               -- &       $24.339$ &     $8.830$ &  $-6.317$ & $103.462$ &                       -- &              $213.858$ \\
      TM1             &    $511.198$ &    $783.000$ &  $770.000$ &        $100.580$ &        $159.111$ &               -- &       $21.459$ &     $7.249$ &   $0.610$ &  $71.308$ &                       -- &                     -- \\
      \bottomrule
    \end{tabular}
  }
\end{table}

%% file: Tables/Table-matter.tex
\begin{table}[t!]
  \centering
  \caption{\label{tab:matter}
    Properties of symmetric nuclear matter at $\rho_{0}$ for various effective interactions.
    Here, $E_{0}$ denotes the binding energy per nucleon.
    The bulk properties are given by coefficients in the power-series expansion of isospin-asymmetric nuclear EoS around $\rho_{0}$ in Sec.~\ref{sec:bulk}.
    The $K_{\rm asy}$, $K_{\rm sat,2}$, and $K_{N}$ are respectively expressed as $K_{\rm asy}=K_{\rm sym}-6L$, $K_{\rm sat,2}(=K_{\tau})=K_{\rm asy}-J_{0}L/K_{0}$, and $K_{N}=K_{\rm sym}+K_{0}$~\citep{Sotani:2022zvu,Lattimer:2023rpe}.}
  \resizebox{\textwidth}{!}{
    \begin{tabular}{llcrrrrrrrrrr}
      \toprule
      \               &  $\rho_{0}$ & $M_{N}^{\ast}$/$M_{N}$ &  $E_{0}$ &  $K_{0}$ &     $J_{0}$ & $E_{\rm sym}$ &      $L$ & $K_{\rm sym}$ & $J_{\rm sym}$ & $K_{\rm asy}$ & $K_{\rm sat,2}$ &   $K_{N}$ \\
      Models          & (fm$^{-3}$) &                      \ &    (MeV) &    (MeV) &       (MeV) &         (MeV) &    (MeV) &         (MeV) &         (MeV) &         (MeV) &           (MeV) &     (MeV) \\
      \midrule
      OMEG0           &    $0.1500$ &        $0.640$ & $-16.45$ & $280.00$ &    $-66.98$ &       $34.55$ &  $50.00$ &     $-384.43$ &     $-533.44$ &     $-684.43$ &       $-672.47$ & $-104.43$ \\
      OMEG1           &    $0.1484$ &        $0.620$ & $-16.38$ & $256.00$ &   $-300.62$ &       $35.06$ &  $70.00$ &     $-218.83$ &      $-68.94$ &     $-638.83$ &       $-556.62$ &   $37.17$ \\
      OMEG2           &    $0.1484$ &        $0.620$ & $-16.38$ & $256.00$ &   $-300.56$ &       $33.00$ &  $45.00$ &     $-216.72$ &     $1020.63$ &     $-486.72$ &       $-433.89$ &   $39.28$ \\
      OMEG3           &    $0.1484$ &        $0.620$ & $-16.38$ & $256.00$ &   $-300.28$ &       $30.00$ &  $20.00$ &      $-65.85$ &     $1449.98$ &     $-185.85$ &       $-162.39$ &  $190.15$ \\
      BigApple        &    $0.155$  &        $0.609$ & $-16.34$ & $226.08$ &   $-209.66$ &       $31.32$ &  $40.00$ &       $89.76$ &     $1121.88$ &     $-150.21$ &       $-113.12$ &  $315.84$ \\
      DINOa           &    $0.1522$ &        $0.626$ & $-16.16$ & $210.00$ &   $-365.42$ &       $31.42$ &  $50.00$ &      $504.96$ &     $9316.53$ &      $204.96$ &        $291.97$ &  $714.96$ \\
      DINOb           &    $0.1525$ &        $0.632$ & $-16.21$ & $207.00$ &   $-412.46$ &       $33.07$ &  $70.00$ &      $608.70$ &     $9692.85$ &      $188.70$ &        $328.18$ &  $815.70$ \\
      DINOc           &    $0.1519$ &        $0.632$ & $-16.22$ & $206.00$ &   $-421.37$ &       $34.58$ &  $90.00$ &      $715.40$ &     $9846.87$ &      $175.40$ &        $359.49$ &  $921.40$ \\
      FSU-$\delta$6.2 &    $0.148$  &        $0.610$ & $-16.31$ & $229.20$ &   $-322.12$ &       $32.53$ &  $48.21$ &     $-311.61$ &      $-92.61$ &     $-600.87$ &       $-533.12$ &  $-82.4$ \\
      FSU-$\delta$6.7 &    $0.148$  &        $0.610$ & $-16.31$ & $229.20$ &   $-322.12$ &       $32.75$ &  $53.50$ &     $-227.23$ &      $673.54$ &     $-548.23$ &       $-473.04$ &    $1.97$ \\
      FSUGarnet       &    $0.153$  &        $0.579$ & $-16.23$ & $229.63$ &      $8.47$ &       $30.92$ &  $50.96$ &       $58.31$ &      $149.10$ &     $-247.45$ &       $-249.33$ &  $287.94$ \\
      FSUGarnet+R     &    $0.1527$ &        $0.582$ & $-16.18$ & $228.77$ &    $-24.37$ &       $30.89$ &  $55.79$ &       $20.00$ &      $417.64$ &     $-314.74$ &       $-308.80$ &  $248.77$ \\
      FSUGold         &    $0.1484$ &        $0.610$ & $-16.30$ & $230.00$ &   $-523.39$ &       $32.59$ &  $60.50$ &      $-51.27$ &      $424.15$ &     $-414.27$ &       $-276.59$ &  $178.73$ \\
      FSUGold2        &    $0.1505$ &        $0.593$ & $-16.28$ & $238.00$ &   $-149.49$ &       $37.62$ & $112.80$ &       $25.21$ &     $-165.75$ &     $-651.59$ &       $-580.74$ &  $263.21$ \\
      FSUGold2+R      &    $0.1522$ &        $0.594$ & $-16.22$ & $241.22$ &    $-36.17$ &       $32.03$ &  $57.20$ &       $-6.89$ &      $971.30$ &     $-350.09$ &       $-341.51$ &  $234.33$ \\
      HPNL0           &    $0.1477$ &        $0.603$ & $-16.10$ & $227.38$ &   $-217.51$ &       $34.63$ &  $78.38$ &      $-69.89$ &      $488.76$ &     $-540.18$ &       $-465.20$ &  $157.48$ \\
      HPNL5           &    $0.1486$ &        $0.618$ & $-16.18$ & $230.32$ &   $-320.74$ &       $33.04$ &  $52.29$ &      $-83.86$ &     $1178.98$ &     $-397.60$ &       $-324.78$ &  $146.46$ \\
      IOPB-I          &    $0.149$  &        $0.594$ & $-16.11$ & $222.33$ &   $-109.15$ &       $33.36$ &  $63.70$ &      $-38.98$ &      $873.02$ &     $-421.18$ &       $-389.00$ &  $183.35$ \\
      IU-FSU          &    $0.1547$ &        $0.609$ & $-16.40$ & $231.30$ &   $-289.40$ &       $31.30$ &  $47.22$ &       $28.61$ &      $370.02$ &     $-254.73$ &       $-195.64$ &  $259.91$ \\
      NL3             &    $0.1483$ &        $0.595$ & $-16.26$ & $271.50$ &    $201.62$ &       $37.28$ & $118.18$ &      $101.01$ &      $181.90$ &     $-608.09$ &       $-695.86$ &  $372.51$ \\
      PD15            &    $0.1484$ &        $0.618$ & $-16.62$ & $229.49$ &   $-399.42$ &       $35.29$ &  $42.60$ &      $-56.52$ &     $1582.27$ &     $-312.10$ &       $-237.96$ &  $172.97$ \\
      TAMUC-FSUa      &    $0.149$  &        $0.601$ & $-16.23$ & $245.31$ &   $-159.85$ &       $35.05$ &  $82.71$ &      $-68.69$ &      $382.54$ &     $-564.96$ &       $-511.07$ &  $176.62$ \\
      TM1             &    $0.145$  &        $0.635$ & $-16.23$ & $280.39$ &   $-286.69$ &       $36.84$ & $110.60$ &       $33.50$ &       $-65.26$ &    $-630.08$ &       $-517.00$ &  $313.89$ \\
      \bottomrule
    \end{tabular}
  }
\end{table}

%% file: Tables/Table-CCs-delta.tex
\begin{table}[t!]
  \centering
  \caption{\label{tab:CCs2}
    Model parameters and nuclear properties for the extended version of the FSUGarnet, TAMUC-FSUa, and FSUGold2 models.
    For each series, the parameter $c_{3}$ is fixed by the original value shown in Table.~\ref{tab:CCs}.}
  \resizebox{\textwidth}{5.45cm}{
  % \resizebox{\textwidth}{!}{
  % \resizebox{17.1cm}{!}{
    \begin{tabular}{lrrrrrrrrrrr}
      \toprule
      \           & $m_{\sigma}$ & $g_{\sigma}^{2}$ & $g_{\omega}^{2}$ & $g_{\delta}^{2}$ & $g_{\rho}^{2}$ & $g_{2}$     &  $g_{3}$ & $\Lambda_{\omega\rho}$ &  $E_{0}$ & $E_{\rm sym}$ & $K_{\rm sym}$ \\
      Models      &        (MeV) &                \ &                \ &                \ &              \ & (fm$^{-1}$) &        \ &                      \ &    (MeV) &         (MeV) &         (MeV) \\
      \midrule
      \multicolumn{12}{c}{FSUGarnet series ($L=50.96$ MeV)} \\
      \midrule
      $\delta$000 &     $496.94$ &         $110.35$ &         $187.69$ &           $0.00$ &        $47.97$ &      $9.56$ &  $-7.12$ &              $1555.73$ & $-16.23$ &       $30.92$ &       $58.31$ \\
      $\delta$050 &     $496.23$ &         $110.06$ &         $187.68$ &          $50.00$ &        $64.41$ &      $9.54$ &  $-7.11$ &               $364.29$ & $-16.25$ &       $32.54$ &       $39.53$ \\
      $\delta$100 &     $496.26$ &         $110.07$ &         $187.68$ &         $100.00$ &        $95.83$ &      $9.54$ &  $-7.11$ &               $290.10$ & $-16.24$ &       $32.50$ &      $162.95$ \\
      $\delta$150 &     $496.41$ &         $110.12$ &         $187.69$ &         $150.00$ &       $128.83$ &      $9.54$ &  $-7.10$ &               $274.42$ & $-16.24$ &       $32.22$ &      $337.44$ \\
      $\delta$200 &     $496.51$ &         $110.16$ &         $187.69$ &         $200.00$ &       $163.34$ &      $9.54$ &  $-7.10$ &               $276.62$ & $-16.23$ &       $31.98$ &      $548.39$ \\
      $\delta$250 &     $496.10$ &         $109.99$ &         $187.69$ &         $250.00$ &       $199.38$ &      $9.53$ &  $-7.10$ &               $286.98$ & $-16.24$ &       $31.85$ &      $790.04$ \\
      $\delta$300 &     $493.90$ &         $109.10$ &         $187.65$ &         $300.00$ &       $238.62$ &      $9.45$ &  $-7.07$ &               $303.57$ & $-16.30$ &       $32.60$ &     $1054.19$ \\
      \midrule
      \multicolumn{12}{c}{TAMUC-FSUa series ($L=82.71$ MeV)} \\
      \midrule
      $\delta$000 &     $502.20$ &         $106.50$ &         $176.18$ &           $0.00$ &        $24.40$ &      $8.83$ &  $-6.32$ &               $213.86$ & $-16.23$ &       $35.05$ &      $-68.69$ \\
      $\delta$050 &     $502.85$ &         $106.74$ &         $176.19$ &          $50.00$ &        $53.83$ &      $8.84$ &  $-6.30$ &               $158.60$ & $-16.21$ &       $34.55$ &        $3.69$ \\
      $\delta$100 &     $502.75$ &         $106.70$ &         $176.19$ &         $100.00$ &        $85.18$ &      $8.83$ &  $-6.30$ &               $159.75$ & $-16.21$ &       $34.55$ &      $104.58$ \\
      $\delta$150 &     $502.75$ &         $106.70$ &         $176.19$ &         $150.00$ &       $117.17$ &      $8.83$ &  $-6.30$ &               $168.70$ & $-16.21$ &       $34.50$ &      $233.64$ \\
      $\delta$200 &     $502.75$ &         $106.70$ &         $176.19$ &         $200.00$ &       $149.71$ &      $8.83$ &  $-6.29$ &               $180.42$ & $-16.21$ &       $34.35$ &      $389.08$ \\
      $\delta$250 &     $502.70$ &         $106.68$ &         $176.19$ &         $250.00$ &       $182.90$ &      $8.83$ &  $-6.28$ &               $193.75$ & $-16.20$ &       $34.15$ &      $568.48$ \\
      $\delta$300 &     $502.10$ &         $106.44$ &         $176.19$ &         $300.00$ &       $217.36$ &      $8.81$ &  $-6.28$ &               $208.75$ & $-16.22$ &       $34.20$ &      $767.29$ \\
      \midrule
      \multicolumn{12}{c}{FSUGold2 series ($L=112.80$ MeV)} \\
      \midrule
      $\delta$000 &     $497.48$ &         $108.07$ &         $183.73$ &           $0.00$ &        $20.15$ &      $8.55$ &  $-1.02$ &                $12.29$ & $-16.28$ &       $37.62$ &       $25.21$ \\
      $\delta$050 &     $498.00$ &         $108.28$ &         $183.74$ &          $50.00$ &        $50.47$ &      $8.56$ &  $-1.00$ &                $77.62$ & $-16.27$ &       $37.55$ &       $87.38$ \\
      $\delta$100 &     $497.96$ &         $108.26$ &         $183.74$ &         $100.00$ &        $81.86$ &      $8.56$ &  $-1.00$ &               $109.82$ & $-16.27$ &       $37.51$ &      $187.87$ \\
      $\delta$150 &     $499.99$ &         $108.27$ &         $183.74$ &         $150.00$ &       $113.85$ &      $8.56$ &  $-1.00$ &               $133.90$ & $-16.27$ &       $37.34$ &      $320.60$ \\
      $\delta$200 &     $498.03$ &         $108.29$ &         $183.74$ &         $200.00$ &       $146.62$ &      $8.56$ &  $-1.00$ &               $155.39$ & $-16.27$ &       $37.12$ &      $481.08$ \\
      $\delta$250 &     $498.09$ &         $108.31$ &         $183.74$ &         $250.00$ &       $179.93$ &      $8.56$ &  $-0.99$ &               $175.51$ & $-16.26$ &       $36.73$ &      $668.45$ \\
      $\delta$300 &     $497.60$ &         $108.12$ &         $183.74$ &         $300.00$ &       $214.65$ &      $8.54$ &  $-1.01$ &               $196.02$ & $-16.27$ &       $36.59$ &      $876.59$ \\
      \bottomrule
    \end{tabular}
  }
\end{table}

%% file: Tables/Table-CCs-rho.tex
\begin{table}[t!]
  \centering
  \caption{\label{tab:CCs3}
    Model parameters and several properties for the FSUGold2 with the $\delta$-$N$ coupling and the quartic self-interaction of $\rho$ meson.
    We set $c_{3}=144.12$ and $g_{\delta}^{2}=300.00$.}
  \resizebox{\textwidth}{!}{
    \begin{tabular}{lrrrrrrrrrrr}
      \toprule
      \         & $m_{\sigma}$ & $g_{\sigma}^{2}$ & $g_{\omega}^{2}$ & $g_{\rho}^{2}$ & $g_{2}$     &  $g_{3}$ &  $e_{3}$ & $\Lambda_{\omega\rho}$ &  $E_{0}$ & $E_{\rm sym}$ & $K_{\rm sym}$ \\
      Models    &        (MeV) &                \ &                \ &              \ & (fm$^{-1}$) &        \ &        \ &                      \ &    (MeV) &         (MeV) &         (MeV) \\
      \midrule
      $\rho$000 &     $497.60$ &         $108.12$ &         $183.74$ &       $214.65$ &      $8.54$ &  $-1.01$ &   $0.00$ &               $196.02$ & $-16.27$ &       $36.59$ &      $876.59$ \\
      $\rho$100 &     $497.60$ &         $108.11$ &         $183.74$ &       $214.97$ &      $8.54$ &  $-1.01$ & $100.00$ &               $196.39$ & $-16.27$ &       $36.75$ &      $875.10$ \\
      $\rho$200 &     $497.62$ &         $108.12$ &         $183.74$ &       $215.17$ &      $8.55$ &  $-1.01$ & $200.00$ &               $196.61$ & $-16.27$ &       $36.85$ &      $874.16$ \\
      $\rho$300 &     $497.64$ &         $108.13$ &         $183.74$ &       $215.39$ &      $8.55$ &  $-1.01$ & $300.00$ &               $196.86$ & $-16.27$ &       $36.96$ &      $873.14$ \\
      $\rho$400 &     $497.65$ &         $108.13$ &         $183.74$ &       $215.73$ &      $8.55$ &  $-1.01$ & $400.00$ &               $197.24$ & $-16.27$ &       $37.13$ &      $871.55$ \\
      $\rho$500 &     $497.71$ &         $108.16$ &         $183.74$ &       $215.95$ &      $8.55$ &  $-1.01$ & $500.00$ &               $197.49$ & $-16.27$ &       $37.24$ &      $870.53$ \\
      $\rho$600 &     $497.83$ &         $108.20$ &         $183.74$ &       $216.11$ &      $8.55$ &  $-1.00$ & $600.00$ &               $197.67$ & $-16.27$ &       $37.32$ &      $869.79$ \\
      $\rho$700 &     $497.79$ &         $108.19$ &         $183.74$ &       $216.51$ &      $8.55$ &  $-1.00$ & $700.00$ &               $198.12$ & $-16.27$ &       $37.52$ &      $867.93$ \\
      $\rho$800 &     $497.83$ &         $108.21$ &         $183.74$ &       $216.91$ &      $8.55$ &  $-1.00$ & $800.00$ &               $198.57$ & $-16.27$ &       $37.72$ &      $866.08$ \\
      \bottomrule
    \end{tabular}
  }
\end{table}

%% file: Tables/Table-NST-all.tex
\begin{table}[p!]
  \centering
  \caption{\label{tab:NST}
    Predictions for the charge radius, $R_{\rm ch}$, neutron skin thickness, $R_{\rm skin}$, weak radius, $R_{\rm wk}$, and weak skin thickness, $R_{\rm wk}-R_{\rm ch}$, of $^{48}$Ca and $^{208}$Pb in fm.
    The PREX-2 and CREX results are also presented.%~\citep{PREX:2021umo,CREX:2022kgg}.
  }
  % \resizebox{15.7cm}{!}{
  % \resizebox{\textwidth}{11.0cm}{
  \resizebox{\textwidth}{!}{
  \begin{tabular}{lccccccccc}
    \toprule
    \               & \multicolumn{4}{c}{$^{48}$Ca} & \ & \multicolumn{4}{c}{$^{208}$Pb} \\
    \cline{2-5}\cline{7-10} \\[-0.30cm]
    Models          & $R_{\rm ch}$ & $R_{\rm skin}$ & $R_{\rm wk}$ & $R_{\rm wk}-R_{\rm ch}$ & \hspace{0.5cm} & $R_{\rm ch}$ & $R_{\rm skin}$ & $R_{\rm wk}$ & $R_{\rm wk}-R_{\rm ch}$ \\
    \midrule
    OMEG0           & $3.499$ & $0.201$ & $3.689$ & $0.189$ & \ & $5.514$ & $0.227$ & $5.738$ & $0.223$ \\
    OMEG1           & $3.495$ & $0.209$ & $3.693$ & $0.198$ & \ & $5.532$ & $0.245$ & $5.774$ & $0.242$ \\
    OMEG2           & $3.500$ & $0.194$ & $3.682$ & $0.183$ & \ & $5.536$ & $0.209$ & $5.741$ & $0.205$ \\
    OMEG3           & $3.510$ & $0.161$ & $3.659$ & $0.149$ & \ & $5.549$ & $0.143$ & $5.685$ & $0.136$ \\
    BigApple        & $3.496$ & $0.168$ & $3.652$ & $0.156$ & \ & $5.513$ & $0.150$ & $5.657$ & $0.144$ \\
    DINOa           & $3.510$ & $0.100$ & $3.597$ & $0.086$ & \ & $5.531$ & $0.175$ & $5.700$ & $0.169$ \\
    DINOb           & $3.514$ & $0.105$ & $3.606$ & $0.092$ & \ & $5.531$ & $0.200$ & $5.726$ & $0.195$ \\
    DINOc           & $3.519$ & $0.113$ & $3.619$ & $0.100$ & \ & $5.532$ & $0.224$ & $5.752$ & $0.220$ \\
    FSU-$\delta$6.2 & $3.501$ & $0.195$ & $3.685$ & $0.183$ & \ & $5.547$ & $0.223$ & $5.766$ & $0.219$ \\
    FSU-$\delta$6.7 & $3.502$ & $0.195$ & $3.685$ & $0.184$ & \ & $5.549$ & $0.225$ & $5.770$ & $0.221$ \\
    FSUGarnet       & $3.471$ & $0.167$ & $3.626$ & $0.155$ & \ & $5.514$ & $0.162$ & $5.670$ & $0.156$ \\
    FSUGarnet+R     & $3.473$ & $0.180$ & $3.643$ & $0.168$ & \ & $5.513$ & $0.184$ & $5.692$ & $0.179$ \\
    FSUGold         & $3.479$ & $0.197$ & $3.664$ & $0.186$ & \ & $5.540$ & $0.207$ & $5.742$ & $0.203$ \\
    FSUGold2        & $3.469$ & $0.232$ & $3.690$ & $0.221$ & \ & $5.512$ & $0.286$ & $5.797$ & $0.285$ \\
    FSUGold2+R      & $3.478$ & $0.188$ & $3.654$ & $0.176$ & \ & $5.509$ & $0.198$ & $5.702$ & $0.193$ \\
    HPNL0           & $3.490$ & $0.213$ & $3.692$ & $0.203$ & \ & $5.551$ & $0.247$ & $5.795$ & $0.244$ \\
    HPNL5           & $3.493$ & $0.194$ & $3.676$ & $0.183$ & \ & $5.545$ & $0.207$ & $5.747$ & $0.202$ \\
    IOPB-I          & $3.493$ & $0.200$ & $3.682$ & $0.189$ & \ & $5.548$ & $0.221$ & $5.765$ & $0.217$ \\
    IU-FSU          & $3.473$ & $0.173$ & $3.634$ & $0.161$ & \ & $5.501$ & $0.161$ & $5.656$ & $0.156$ \\
    NL3             & $3.490$ & $0.226$ & $3.705$ & $0.215$ & \ & $5.529$ & $0.280$ & $5.807$ & $0.278$ \\
    PD15            & $3.520$ & $0.197$ & $3.706$ & $0.186$ & \ & $5.561$ & $0.205$ & $5.762$ & $0.201$ \\
    TAMUC-FSUa      & $3.483$ & $0.215$ & $3.687$ & $0.204$ & \ & $5.528$ & $0.250$ & $5.776$ & $0.248$ \\
    TM1             & $3.499$ & $0.227$ & $3.715$ & $0.216$ & \ & $5.556$ & $0.271$ & $5.825$ & $0.269$ \\
    \midrule
    \multicolumn{10}{c}{FSUGarnet series} \\
    \midrule
    $\delta$000     & $3.471$ & $0.167$ & $3.626$ & $0.155$ & \ & $5.514$ & $0.162$ & $5.670$ & $0.156$ \\
    $\delta$050     & $3.466$ & $0.176$ & $3.630$ & $0.164$ & \ & $5.496$ & $0.200$ & $5.691$ & $0.195$ \\
    $\delta$100     & $3.469$ & $0.160$ & $3.616$ & $0.148$ & \ & $5.497$ & $0.193$ & $5.685$ & $0.188$ \\
    $\delta$150     & $3.472$ & $0.138$ & $3.597$ & $0.125$ & \ & $5.480$ & $0.181$ & $5.675$ & $0.175$ \\
    $\delta$200     & $3.477$ & $0.110$ & $3.573$ & $0.097$ & \ & $5.504$ & $0.166$ & $5.664$ & $0.160$ \\
    $\delta$250     & $3.481$ & $0.078$ & $3.544$ & $0.063$ & \ & $5.508$ & $0.151$ & $5.652$ & $0.144$ \\
    $\delta$300     & $3.470$ & $0.050$ & $3.505$ & $0.035$ & \ & $5.509$ & $0.142$ & $5.645$ & $0.136$ \\
    \midrule
    \multicolumn{10}{c}{TAMUC-FSUa series} \\
    \midrule
    $\delta$000     & $3.483$ & $0.215$ & $3.687$ & $0.204$ & \ & $5.528$ & $0.250$ & $5.776$ & $0.248$ \\
    $\delta$050     & $3.484$ & $0.203$ & $3.676$ & $0.192$ & \ & $5.528$ & $0.245$ & $5.770$ & $0.242$ \\
    $\delta$100     & $3.487$ & $0.190$ & $3.666$ & $0.179$ & \ & $5.529$ & $0.240$ & $5.765$ & $0.237$ \\
    $\delta$150     & $3.491$ & $0.174$ & $3.653$ & $0.162$ & \ & $5.530$ & $0.232$ & $5.759$ & $0.229$ \\
    $\delta$200     & $3.495$ & $0.154$ & $3.637$ & $0.142$ & \ & $5.532$ & $0.223$ & $5.751$ & $0.219$ \\
    $\delta$250     & $3.501$ & $0.129$ & $3.617$ & $0.116$ & \ & $5.536$ & $0.211$ & $5.742$ & $0.207$ \\
    $\delta$300     & $3.506$ & $0.101$ & $3.593$ & $0.087$ & \ & $5.538$ & $0.200$ & $5.734$ & $0.195$ \\
    \midrule
    \multicolumn{10}{c}{FSUGold2 series} \\
    \midrule
    $\delta$000     & $3.469$ & $0.232$ & $3.690$ & $0.221$ & \ & $5.512$ & $0.286$ & $5.797$ & $0.285$ \\
    $\delta$050     & $3.470$ & $0.222$ & $3.681$ & $0.211$ & \ & $5.511$ & $0.284$ & $5.793$ & $0.282$ \\
    $\delta$100     & $3.473$ & $0.210$ & $3.672$ & $0.199$ & \ & $5.511$ & $0.278$ & $5.788$ & $0.277$ \\
    $\delta$150     & $3.477$ & $0.194$ & $3.660$ & $0.183$ & \ & $5.513$ & $0.270$ & $5.781$ & $0.268$ \\
    $\delta$200     & $3.482$ & $0.175$ & $3.645$ & $0.163$ & \ & $5.516$ & $0.258$ & $5.772$ & $0.256$ \\
    $\delta$250     & $3.487$ & $0.150$ & $3.625$ & $0.138$ & \ & $5.521$ & $0.244$ & $5.762$ & $0.241$ \\
    $\delta$300     & $3.494$ & $0.123$ & $3.603$ & $0.109$ & \ & $5.525$ & $0.230$ & $5.753$ & $0.226$ \\
    \midrule
    Experiment\hspace{1.8cm} & $3.477$ & $0.121$ & $3.636$ & $0.159$ & \ & $5.501$ & $0.283$ & $5.800$ & $0.299$ \\
    \bottomrule
  \end{tabular}
  }
\end{table}

%% file: Tables/Table-NS-all.tex
\begin{table}[p!]
  \centering
  \caption{\label{tab:NS}Properties of neutron stars.
    We calculate the crust-core transition density, $\rho_{t}$, and the observables at the canonical- and maximum-mass points.}
  % \resizebox{\textwidth}{11.5cm}{
  \resizebox{\textwidth}{!}{
  \tabcolsep=0.3cm
  \begin{tabular}{lcccrcccr}
    \toprule
    \               &  $\rho_{t}$ & $R_{1.4}$ & $\rho_{1.4}$ & $\Lambda_{1.4}$ & $M_{\rm max}$ & $R_{\rm max}$ & $\rho_{\rm max}$ & $\Lambda_{\rm max}$ \\
    Models          & (fm$^{-1}$) &      (km) &  (fm$^{-1}$) &               \ & ($M_{\odot}$) &          (km) &      (fm$^{-1}$) &                   \ \\
    \midrule
    OMEG0           &     $0.093$ &   $12.43$ &      $0.359$ &           $498$ &        $2.61$ &       $12.16$ &          $0.772$ &                 $4$ \\
    OMEG1           &     $0.079$ &   $12.76$ &      $0.402$ &           $515$ &        $2.13$ &       $11.70$ &          $0.917$ &                $14$ \\
    OMEG2           &     $0.093$ &   $12.40$ &      $0.428$ &           $458$ &        $2.07$ &       $11.31$ &          $0.980$ &                $13$ \\
    OMEG3           &     $0.104$ &   $12.40$ &      $0.409$ &           $462$ &        $2.07$ &       $11.39$ &          $0.962$ &                $14$ \\
    BigApple        &     $0.096$ &   $13.03$ &      $0.329$ &           $677$ &        $2.60$ &       $12.38$ &          $0.759$ &                 $5$ \\
    DINOa           &     $0.092$ &   $14.11$ &      $0.307$ &          $1047$ &        $2.17$ &       $12.30$ &          $0.854$ &                $16$ \\
    DINOb           &     $0.087$ &   $14.37$ &      $0.306$ &          $1136$ &        $2.15$ &       $12.26$ &          $0.870$ &                $16$ \\
    DINOc           &     $0.083$ &   $14.63$ &      $0.301$ &          $1237$ &        $2.15$ &       $12.30$ &          $0.871$ &                $16$ \\
    FSU-$\delta$6.2 &     $0.083$ &   $12.08$ &      $0.409$ &           $416$ &        $2.10$ &       $11.53$ &          $0.922$ &                $16$ \\
    FSU-$\delta$6.7 &     $0.082$ &   $12.82$ &      $0.386$ &           $573$ &        $2.05$ &       $11.76$ &          $0.917$ &                $20$ \\
    FSUGarnet       &     $0.084$ &   $12.87$ &      $0.382$ &           $596$ &        $2.07$ &       $11.66$ &          $0.932$ &                $17$ \\
    FSUGarnet+R     &     $0.083$ &   $12.94$ &      $0.384$ &           $616$ &        $2.06$ &       $11.66$ &          $0.938$ &                $18$ \\
    FSUGold         &     $0.079$ &   $12.32$ &      $0.520$ &           $400$ &        $1.72$ &       $10.80$ &          $1.156$ &                $32$ \\
    FSUGold2        &     $0.061$ &   $14.00$ &      $0.351$ &           $873$ &        $2.07$ &       $12.09$ &          $0.904$ &                $19$ \\
    FSUGold2+R      &     $0.087$ &   $13.19$ &      $0.347$ &           $715$ &        $2.26$ &       $12.10$ &          $0.848$ &                $12$ \\
    HPNL0           &     $0.072$ &   $13.32$ &      $0.383$ &           $674$ &        $2.03$ &       $11.74$ &          $0.946$ &                $19$ \\
    HPNL5           &     $0.088$ &   $12.76$ &      $0.408$ &           $551$ &        $2.01$ &       $11.43$ &          $0.980$ &                $18$ \\
    IOPB-I          &     $0.081$ &   $13.21$ &      $0.362$ &           $693$ &        $2.15$ &       $11.92$ &          $0.890$ &                $15$ \\
    IU-FSU          &     $0.088$ &   $12.49$ &      $0.433$ &           $482$ &        $1.94$ &       $11.19$ &          $1.027$ &                $19$ \\
    NL3             &     $0.063$ &   $14.68$ &      $0.272$ &          $1255$ &        $2.77$ &       $13.30$ &          $0.668$ &                 $4$ \\
    PD15            &     $0.094$ &   $12.49$ &      $0.434$ &           $488$ &        $1.92$ &       $11.19$ &          $1.026$ &                $21$ \\
    TAMUC-FSUa\hspace{1.0cm}      &     $0.073$ &   $13.48$ &      $0.365$ &           $725$ &        $2.10$ &       $11.92$ &          $0.909$ &                $17$ \\
    TM1             &     $0.068$ &   $14.31$ &      $0.320$ &          $1038$ &        $2.18$ &       $12.37$ &          $0.852$ &                $16$ \\
    \midrule
    \multicolumn{9}{c}{FSUGarnet series} \\
    \midrule
    $\delta$000     &     $0.084$ &   $12.87$ &      $0.382$ &           $596$ &        $2.07$ &       $11.66$ &          $0.932$ &                $17$ \\
    $\delta$050     &     $0.094$ &   $13.09$ &      $0.360$ &           $681$ &        $2.10$ &       $11.91$ &          $0.893$ &                $18$ \\
    $\delta$100     &     $0.097$ &   $13.37$ &      $0.336$ &           $784$ &        $2.13$ &       $12.15$ &          $0.859$ &                $19$ \\
    $\delta$150     &     $0.097$ &   $13.62$ &      $0.318$ &           $893$ &        $2.14$ &       $12.31$ &          $0.841$ &                $20$ \\
    $\delta$200     &     $0.097$ &   $13.82$ &      $0.307$ &           $973$ &        $2.13$ &       $12.40$ &          $0.833$ &                $21$ \\
    $\delta$250     &     $0.096$ &   $13.96$ &      $0.301$ &          $1031$ &        $2.13$ &       $12.42$ &          $0.834$ &                $22$ \\
    $\delta$300     &     $0.095$ &   $14.02$ &      $0.300$ &          $1059$ &        $2.12$ &       $12.39$ &          $0.840$ &                $22$ \\
    \midrule
    \multicolumn{9}{c}{TAMUC-FSUa series} \\
    \midrule
    $\delta$000     &     $0.073$ &   $13.48$ &      $0.365$ &           $725$ &        $2.10$ &       $11.92$ &          $0.909$ &                $17$ \\
    $\delta$050     &     $0.076$ &   $13.72$ &      $0.338$ &           $850$ &        $2.14$ &       $12.19$ &          $0.867$ &                $17$ \\
    $\delta$100     &     $0.080$ &   $13.94$ &      $0.318$ &           $947$ &        $2.16$ &       $12.39$ &          $0.839$ &                $18$ \\
    $\delta$150     &     $0.083$ &   $14.15$ &      $0.302$ &          $1073$ &        $2.17$ &       $12.53$ &          $0.824$ &                $19$ \\
    $\delta$200     &     $0.085$ &   $14.33$ &      $0.292$ &          $1158$ &        $2.17$ &       $12.60$ &          $0.818$ &                $20$ \\
    $\delta$250     &     $0.086$ &   $14.48$ &      $0.285$ &          $1249$ &        $2.16$ &       $12.63$ &          $0.817$ &                $21$ \\
    $\delta$300     &     $0.086$ &   $14.58$ &      $0.283$ &          $1276$ &        $2.16$ &       $12.63$ &          $0.820$ &                $21$ \\
    \midrule
    \multicolumn{9}{c}{FSUGold2 series} \\
    \midrule
    $\delta$000     &     $0.061$ &   $14.00$ &      $0.351$ &           $873$ &        $2.07$ &       $12.09$ &          $0.904$ &                $19$ \\
    $\delta$050     &     $0.064$ &   $14.13$ &      $0.335$ &           $950$ &        $2.08$ &       $12.25$ &          $0.881$ &                $21$ \\
    $\delta$100     &     $0.069$ &   $14.29$ &      $0.318$ &          $1045$ &        $2.09$ &       $12.39$ &          $0.861$ &                $22$ \\
    $\delta$150     &     $0.074$ &   $14.45$ &      $0.304$ &          $1151$ &        $2.10$ &       $12.49$ &          $0.849$ &                $23$ \\
    $\delta$200     &     $0.078$ &   $14.58$ &      $0.295$ &          $1227$ &        $2.10$ &       $12.54$ &          $0.844$ &                $24$ \\
    $\delta$250     &     $0.080$ &   $14.68$ &      $0.289$ &          $1297$ &        $2.09$ &       $12.56$ &          $0.843$ &                $24$ \\
    $\delta$300     &     $0.082$ &   $14.74$ &      $0.287$ &          $1334$ &        $2.08$ &       $12.55$ &          $0.846$ &                $24$ \\
    \bottomrule
  \end{tabular}}
\end{table}